\documentclass[lettersize,journal]{IEEEtran}
\usepackage{amsmath,amsfonts,amssymb}
\usepackage{algorithmic}
\usepackage{algorithm}
\usepackage{array}
\usepackage{eurosym}
\usepackage{diagbox}
\usepackage{bm}
\usepackage{gensymb}
\usepackage{caption}
\usepackage{float}
\usepackage{textcomp}
\usepackage{stfloats}
\usepackage{url}
\usepackage{enumerate}
\usepackage{nccmath} 
\usepackage{verbatim}
\usepackage{graphicx}
\usepackage{cite}
\usepackage{graphicx}
\usepackage{textcomp}
\usepackage{xcolor}
\usepackage{bm}
\usepackage{subcaption}
\makeatletter \renewcommand{\maketag@@@}[1]{\hbox{\m@th\normalsize\normalfont#1}}% \makeatother
\usepackage{subfloat}
\usepackage{balance}
\usepackage{autobreak}
\usepackage{url}
\usepackage{cite}
\usepackage {epstopdf}
\usepackage{multirow}
\allowdisplaybreaks[4]
\usepackage{orcidlink} %
\hypersetup{hidelinks} % hide the hyperlink box

\begin{document}
	\title{Movable Antenna Empowered Covert Dual-Functional Radar-Communication}
	
	\author{Ran Yang,  Ning Wei, \textit{Member, IEEE,} Zheng Dong, \textit{Member, IEEE,} Lin Zhang, \textit{Member, IEEE,}\\  Wanting Lyu,  Yue Xiu, \textit{Member, IEEE,} Ahmad Bazzi, \textit{Senior Member, IEEE,} and Chadi Assi, \textit{Fellow, IEEE}\vspace{-5mm}
	% <-this % stops a space
			\thanks{A preliminary version of this work is available in~\cite{2025arXiv251009949Y}. 
				
				Ran Yang, Ning Wei, Lin Zhang, Wanting Lyu, and Yue Xiu are with the National Key Laboratory of Wireless Communications, University of Electronic Science and Technology of China, Chengdu 611731, China (e-mail: yangran6710@outlook.com; wn@uestc.edu.cn; linzhang1913@uestc.edu.cn; lyuwanting@yeah.net; xiuyue12345678@163.com). 
					
				Zheng Dong is with the School of Information Science and Engineering, Shandong University, Qingdao 266237, China (e-mail:  zhengdong@sdu.edu.cn).

				Ahmad Bazzi is with the Engineering Division, New York University (NYU)
				Abu Dhabi, Abu Dhabi, United Arab Emirates, and NYU WIRELESS,
				NYU Tandon School of Engineering, Brooklyn, 11201, NY, USA  (e-mail:
				ahmad.bazzi@nyu.edu).

			 Chadi Assi is with Concordia University, Montreal, Quebec, H3G 1M8,
			Canada (Email:assi@ciise.concordia.ca).

	}}% <-this % stops a space

	%	\markboth{Journal of \LaTeX\ Class Files,~Vol.~14, No.~8, August~2021}%
	%	{Shell \MakeLowercase{\textit{et al.}}: A Sample Article Using IEEEtran.cls for IEEE Journals}
	
	%	\IEEEpubid{0000--0000/00\$00.00~\copyright~2021 IEEE}
	
	\maketitle
%	\thispagestyle{empty}
	%	\vspace{-10mm}
	\begin{abstract}
		Movable antenna (MA) has emerged as a promising technology to flexibly reconfigure wireless channels by adjusting antenna placement. In this paper, we study a secured dual-functional radar-communication (DFRC) system aided by movable antennas. To enhance the  communication security, we aim to maximize the achievable sum rate by jointly optimizing the {transmitter}  beamforming vectors, receiving filter, and antenna placement, subject to radar signal-to-noise ratio (SINR) and transmission covertness constraints. We consider multiple Willies operating in both non-colluding and colluding modes. For non-colluding Willies, we first employ a Lagrangian dual transformation procedure to reformulate the challenging optimization problem into a more tractable form. Subsequently, we develop an efficient block coordinate descent (BCD) algorithm that integrates semidefinite relaxation (SDR), projected gradient descent (PGD), Dinkelbach transformation, and successive convex approximation (SCA) techniques to tackle the resulting problem. For colluding Willies, we first derive the minimum detection error probability (DEP) by characterizing the optimal detection statistic, which is proven to follow the generalized Erlang distribution. Then, we develop a minimum mean square error (MMSE)-based algorithm to address the {colluding detection problem.} We further provide a comprehensive complexity analysis on the unified design framework. Simulation results demonstrate that the proposed method can significantly improve the covert sum rate, and achieve a {superior balance}  between communication and radar performance compared with existing benchmark schemes.
	\end{abstract}
	
	\vspace{-3.5mm}
	\section{Introduction}
	\IEEEPARstart{T}{he} burgeoning demand for intelligent and ubiquitous services, such as smart grids, unmanned aerial vehicles (UAVs), and the Industrial Internet of Things (IIoT), has become a primary catalyst for the evolution toward sixth-generation (6G) mobile networks. These emerging applications necessitate unprecedented performance metrics, including a peak data rate of 1 Tb/s, a user-experienced data rate of 10–100 Gb/s, and centimetric localization accuracy~\cite{tyagi20256g}. To fulfill these stringent requirements, dual-functional radar-communication (DFRC) has emerged as a pivotal transformative technology within the International Mobile Telecommunication (IMT)-2030 framework~\cite{gonzalez2025six}. In particular, DFRC enables dual functions to share spectrum resources, hardware facilities, and signal-processing modules, leading to significantly improved system capacity and resource utilization efficiency~\cite{10944644}, thanks to the integration of radar sensing and communication functions in a unified platform. {However, the broadcast nature of wireless channels poses significant  security vulnerabilities. Specifically, in  DFRC systems, allowing unified probing waveforms to carry private information will cause a high security risk of being wiretapped if the sensing targets are malicious eavesdroppers~\cite{9755276,su2023security}.}  
	Consequently, the development of efficient security solutions for DFRC systems is highly imperative.

	Recently, various physical layer security (PLS) measures have been  explored for DFRC systems~\cite{10608156}. {Although these existing  approaches, such as artificial noise (AN)~\cite{zou2024securing}, symbol-level precoding (SLP)~\cite{10844869}, and directional modulation (DM)~\cite{chen2025survey} can be employed to protect confidential information from interception, they failed to mitigate the threat to users’ privacy from the discovery of the existence of  confidential message itself~\cite{10090449}.}  Once the transmission behavior is detected, it may lead to the exposure of its location information. In such cases, many cryptographic schemes can be defeated by a determined adversary using non-computational means such as side-channel analysis, thereby leaving networks  with certain security vulnerabilities\cite{7355562}.

	To  fulfill  the demands for  ever-increasing security requirements, covert communication, which shields confidential transmission behaviors from Willies, has been proposed to provide a higher level of security\cite{7447769}. The achievability of the square root law was first established in~\cite{bash2013limits} where Alice can covertly transmit a maximum of $\mathcal{O}(\sqrt{n})$  bits of information to Bob over $n$ channel uses. The authors in~\cite{sobers2017covert} further proposed that $\mathcal{O}(n)$ bits can be covertly transmitted with the aid of AN. To overcome these information-theoretic limitations,  various advanced
	technologies are employed to increase the covert transmission rate in DFRC systems~\cite{hu2024ris,wu2025covert,ma2022covert,jia2024robust,hu2023covert}. For instance, in~\cite{hu2024ris}, a multi-strategy alternating optimization framework was proposed to enhance transmission covertness for reconfigurable intelligent surface (RIS)-enhanced DFRC systems. The authors in~\cite{wu2025covert} studied covert DFRC systems against multiple randomly distributed Willies. A robust beamforming optimization model was studied in~\cite{jia2024robust} to ensure data covertness. 
	Although these works have demonstrated the effectiveness of covert communications, the inherent existence of {wireless fading caused by complex propagation environment} severely undermines covertness performance. Furthermore, existing literature
	mainly focused on the transceiver design by employing conventional fixed-position antennas~(FPAs), while channel variations in the continuous spatial field were not fully exploited. Additionally, the rigid array manifold of FPAs fails to mitigate the spatial conflict between sensing and communication directions, leading to unavoidable array-gain loss and degraded security.
	
	To overcome  the  bottleneck of conventional  FPA-based systems, movable antennas (MAs), or  fluid antennas, have recently been proposed as a promising solution to  enhance the dual-task performance~\cite{zhu2025tutorial,yang2025robust,11353414,11316665,11355857,2025arXiv250813839X}. In MA-assisted systems, each antenna element is connected to a radio frequency (RF) chain via flexible cables to support active antenna movement. A prototype of the MA-assisted radar system was initially demonstrated in~\cite{7360379}, while the latest version achieved an antenna positioning accuracy of up to 0.05 mm in~\cite{11224420}.  Channel modeling and performance analysis were explored  under  far-field channel conditions  in~\cite{zhu2023modeling}. Based on the results in~\cite{zhu2023modeling}, a few works have investigated secure designs for MA-enhanced DFRC systems~\cite{ma2025movable,cao2025joint,le2025beamforming,11223111}. However, the majority of existing literature, i.e., \cite{ma2025movable, cao2025joint, le2025beamforming} and the references therein, primarily focused on safeguarding confidential information from being deciphered instead of  concealing the transmission itself, which deteriorates data covertness. To achieve higher security, the authors in \cite{11223111} pioneered the investigation of covert transmission in MA-enabled DFRC systems. Unfortunately, the proposed approach is limited to scenarios with a single radar target and a solitary Willie. To the best of our knowledge, the potential of MAs for enhancing  transmission covertness within DFRC systems remains largely under-explored, particularly in more complex and practical scenarios. To further bolster security performance, a unified covert design framework is highly desirable.
	
%	Although the current approaches can effectively hide the legitimate message in a chaotic order to prevent eavesdroppers from decoding the confidential information in received signals, they cannot protect the transmission behaviour itself from being detected, which inevitably compromises the credibility of existing secure transmission schemes.

	In this paper, we study a movable antenna-enhanced covert DFRC system against multiple Willies. Both non-colluding and colluding Willies are considered. The optimal detection strategies for Willies are derived, and the corresponding detection error probability (DEP) is analyzed. The main contributions of this paper are summarized as follows.
	\begin{itemize}
		\item We introduce a new {secured}  DFRC system enhanced by movable antennas. To guarantee transmission covertness, we develop a unified design framework  to maximize the achievable  sum rate by jointly designing the transmit beamforming vector, antenna placement, and receiving filter, subject to radar signal-to-interference-plus-noise ratio (SINR) and  {data communication} covertness constraints.	Both non-colluding and colluding Willies are considered.
		\item For non-colluding Willies, we first employ a Lagrangian dual transformation process to reformulate the intractable  optimization problem into a more tractable form. Then, we develop a block coordinate descent (BCD) algorithm that integrates semidefinite relaxation (SDR), projected gradient descent (PGD), Dinkelbach transformation, and successive convex approximation (SCA) techniques to tackle the covert rate maximization problem.
		\item For colluding Willies, we derive the minimum DEP  based on the optimal detection statistic, which is proven to follow the generalized Erlang distribution. Then, based on the Woodbury formula and Pinsker's inequality, we develop a minimum mean square error (MMSE)-based algorithm to address the colluding detection problem.
		\item {We provide a comprehensive complexity analysis of the proposed algorithm, showing that most subproblems can be solved in closed-form (or semi-closed-form), which makes the proposed covert transmission framework highly efficient. }
		\item 	Simulation results demonstrate that the proposed algorithm converges rapidly and significantly enhances the achievable covert rate. Furthermore, the MA-enhanced designs achieve a superior trade-off between radar and communication performance compared to existing benchmarks. Notably, in the presence of colluding Willies, the achievable covert throughput exhibits strong robustness against variations in the required covertness level, particularly when compared to the non-colluding case.
	\end{itemize}

%	\begin{itemize}
%		\item We introduce a new secured DFRC design which maximizes the covert sum rate by jointly designing the trans-
%		mit beamforming vectors, receiving filter, and antenna placement. 
%		\item We develop a block coordinate descent (BCD) algorithm for the covert rate maximization problem, incorporating semidefinite relaxation (SDR), projected gradient descent (PGD), and successive convex approximation (SCA) methods. It is worth noting that many variables can be optimized in closed forms (or semi-closed forms), which makes the proposed algorithm computationally efficient.
%		\item We show that MAs can substantially enhance the covert sum rate, and achieve a satisfactory trade-off between the radar performance and communication quality compared to the baseline schemes.
%	\end{itemize}
	\section{System Model}
	\begin{figure}[t]
		\centering
		\includegraphics[width=0.40\textwidth]{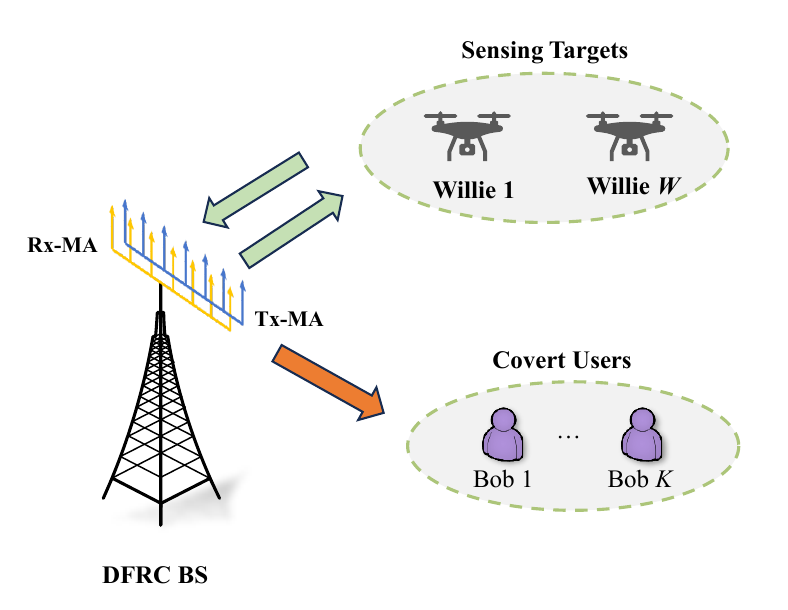}
		\captionsetup{font={normalsize},labelsep=period,singlelinecheck=off}
		\caption{The MA-enhanced DFRC system.} 
		\vspace{-6mm}\label{system}
	\end{figure}%
We consider a narrowband DFRC system, as depicted in Fig.~\ref{system}, where  a dual-functional base station (BS) serves $K$ covert users (Bobs) while simultaneously sensing $W$ point-like targets. The BS is equipped with two separate MA-based uniform linear arrays (ULAs), each consisting of $N$ movable antennas, dedicated to signal transmission and reception, respectively\footnote{We assume that self-interference (SI) between transmit and receive antenna arrays is effectively suppressed by advanced SI cancellation methods, such as physical isolation and digital cancellation~\cite{yang2025robust}.}.  The sensing targets are assumed to be  malicious Willies that attempt  to detect whether the BS is transmitting confidential information or not. We  exploit dedicated radar signals as a cover to achieve covert communications. We assume that the feasible movement range for both the transmitting and receiving MAs is a one-dimensional (1D) interval of length $D$. The transmitting and receiving antenna positioning vectors (APVs) are denoted by $\boldsymbol{t} = [t_{1},t_{2},\dots,t_{N}]^T\in \mathbb{R}^{N\times 1}$ and $\boldsymbol{r} = [r_{1},r_{2},\dots,r_{N}]^T\in \mathbb{R}^{N\times 1}$, respectively, with $0 \le t_1\le t_2 \dots \le t_{N}\le D$ and $0 \le r_1\le r_2 \dots \le r_{N}\le D$.
	\vspace{-2mm}
	\subsection{Signal Model}
	Denote by $\mathcal{H}_0$ and $\mathcal{H}_1$ hypotheses that the BS doesn't transmit covert signals \& does transmit covert signals, respectively. The transmitted signal can be given by
	\begin{equation} \label{hypotheses}
		\begin{cases} 
			\mathcal {H}_{0}:~\boldsymbol{x}(m) = \boldsymbol{r}(m), \;\;\\
			\mathcal {H}_{1}:~\boldsymbol{x}(m) = \sum_{k=1}^{K}\boldsymbol{w}_ks_k(m) + \boldsymbol{r}(m).
		\end{cases}
	\end{equation}
	Here, $\boldsymbol{s}(m) = [s_1(m),s_2(m),\dots,s_K(m)]^T\in \mathbb{C}^{K\times 1}$  denotes the communication symbols for $K$ covert users in the $m$-th time slot, $\forall m\in\mathcal{M} = \{1,\dots,M\}$. Meanwhile,  $\boldsymbol{W} = [\boldsymbol{w}_1,\dots,\boldsymbol{w}_K]\in \mathbb{C}^{N\times K}$ denotes the transmitting beamforming matrix, and $\boldsymbol{r}(m)\in\mathbb{C}^{N\times 1}$ is the dedicated radar signal. Following the work in \cite{10938377,9124713}, we assume that both $\boldsymbol{s}(m)$ and $\boldsymbol{r}(m)$ 
	are zero-mean, temporally-white, and wide-sense stationary stochastic processes, and the symbols are uncorrelated between different users. It is assumed that $\boldsymbol{s}(m)$ and $\boldsymbol{r}(m)$ are independently Gaussian distributed with $\boldsymbol{s}(m)\sim\mathcal{CN}(\boldsymbol{0},\boldsymbol{I}_K)$  and $\boldsymbol{r}(m)\sim\mathcal{CN}(\boldsymbol{0},\boldsymbol{R}_0)$, respectively, where $\boldsymbol{R}_0 \in \mathbb{C}^{N\times N}$ is the covariance matrix of a general rank due to potential multiple beam transmission. We note that once $\boldsymbol{R}_0$ is determined, the dedicated radar signal $\boldsymbol{r}(m)$ can be generated~\cite{4276989}.  Consequently, the covariance matrix of the transmitted signal $\boldsymbol{x}(m)$ can be derived as
	\begin{equation}
		\begin{cases} 
			\mathcal{H}_0: \boldsymbol{R}_{X}^0 =  \boldsymbol{R}_0,  \;\;\\ \mathcal{H}_1: \boldsymbol{R}_{X}^1 = \sum_{k=1}^{K}\boldsymbol{w}_k\boldsymbol{w}_k^H + \boldsymbol{R}_0,
		\end{cases}
	\end{equation} 
	where $\boldsymbol{R}_X^i,\forall i \in \{0,1\}$ denotes the covariance matrix of transmitted signals $\boldsymbol{x}(m)$ under $\mathcal{H}_i$.
	\subsection{Communication Model }
	Given that the signal propagation  distance is significantly larger than the size of moving regions, the far-field response is adopted for channel modeling\cite{zhu2025tutorial}. Specifically, the angle-of-arrival (AoA), angle-of-departure (AoD), and amplitude of the complex coefficient for each link remain constant despite the movement of  MAs. Note that we adopt the geometric model in  \cite{zhu2025tutorial} for communication channels, thus the number of  scattering paths at transceiver nodes is the same\cite{zhu2023modeling}.    Denote by $L_k$ the number of propagation paths between the BS and Bob $k$, where the azimuth angle of the $j$-th path at the BS is given by $\psi_{k}^j\in [0,\pi]$.  Then, the signal propagation difference between the position of the $n$-th transmitting MA $t_n$ and the reference point $o^t$ is given by
	\begin{equation}
		\rho(t_n,\psi_k^j) = t_n\cos\psi_{k}^j,\forall k,j,n.
	\end{equation}
	Consequently, the field response vector (FRV) at $t_n$ can be given by 
	\begin{equation}
		\boldsymbol{g}_k(t_n) = \left[e^{\jmath\frac{2\pi}{\lambda}\rho(t_n,\psi_k^1)},\dots,e^{\jmath\frac{2\pi}{\lambda}\rho(t_n,\psi_k^{L_k})}\right]^T\in \mathbb{C}^{L_k\times 1},
	\end{equation} 
	where $\lambda$ is the carrier wavelength. Therefore,  the field response matrix (FRM) of the link from the BS to Bob $k$ for all $N$ transmitting MAs is given by
		\begin{equation}
		\boldsymbol{G}_k({\boldsymbol{t}})\triangleq\left[\boldsymbol{g}_k({t}_1),\boldsymbol{g}_k({t}_2),\dots,\boldsymbol{g}_k({t}_{N})\right]\in \mathbb{C}^{L_k\times N},
	\end{equation}
	Let $\boldsymbol{\Sigma}_k = \text{diag}\{\sigma_{k,1},\sigma_{k,2},\dots,\sigma_{k,L_k}\}\in\mathbb{C}^{L_k\times L_k}$ denote the path response matrix (PRM), and the channel matrix between the BS and Bob $k$ is given by 
	\begin{equation}
		\boldsymbol{h}_k^H({\boldsymbol{t}}) = \boldsymbol{1}_{L_k}^H\boldsymbol{\Sigma}_k\boldsymbol{G}_k({\boldsymbol{t}})\in \mathbb{C}^{1\times N}, 1\le k \le K,
	\end{equation}
	where the all-ones vector $\boldsymbol{1}_{L_k}\in\mathbb{R}^{L_k\times 1}$ characterizes the FRV associated with the $L_k$ scattering paths to Bob $k$. 
	In the considered system, a quasi-static block fading channel is assumed, and the received signal at the $k$-th user under hypothesis $\mathcal{H}_1$ is given by
	\begin{small}
		\begin{align}
			&y_k(m) = \notag\\ &\underbrace{\boldsymbol{h}_k^H({\boldsymbol{t}})\boldsymbol{w}_ks_k(m)}_{\text{desired signal}}  + \underbrace{\sum_{j\neq k}^{K}\boldsymbol{h}_k^H({\boldsymbol{t}})\boldsymbol{w}_js_j(m)+\boldsymbol{h}_k^H({\boldsymbol{t}})\boldsymbol{r}(m)}_{\text{inter-user interference}} +n_k(m),\notag
		\end{align}
	\end{small}where $n_k(m)\sim \mathcal{CN}(0,\sigma_k^2)$ is the additive white Gaussian noise (AWGN) at Bob $k$. As such, the signal-to-interference-plus-noise (SINR) at Bob $k$ is given by
	\begin{equation}
		\gamma_k = \frac{|\boldsymbol{h}_k^H({\boldsymbol{t}})\boldsymbol{w}_k|^2}{\sum_{j\neq k}^{K}|\boldsymbol{h}_k^H({\boldsymbol{t}})\boldsymbol{w}_j|^2+\boldsymbol{h}_k^H({\boldsymbol{t}})\boldsymbol{R}_0\boldsymbol{h}_k({\boldsymbol{t}})+\sigma_k^2},
	\end{equation}
	and the achievable rate is given by
	$R_k = \log_2(1+\gamma_k),$ which is also known as the covert rate\cite{10090449}.
	\vspace{-2mm}
	
	\subsection{Radar Model}
	We adopt the line-of-sight (LoS) channel model for sensing channels between the BS and targets. Let $\varphi_w$ denote the azimuth angle between  the BS and the $w$-th target, and the receiving and transmitting steering vectors can be given by 
	$\mathbf{a}_r(\varphi_w,{\boldsymbol{r}}) = [e^{\jmath\frac{2\pi}{\lambda}\rho({r}_1,\varphi_w)},\dots,e^{\jmath\frac{2\pi}{\lambda}\rho({r}_{N},\varphi_w)}]^T$  and $\mathbf{a}_t(\varphi_w,{\boldsymbol{t}}) = [e^{\jmath\frac{2\pi}{\lambda}\rho({t}_1,\varphi_w)},\dots,e^{\jmath\frac{2\pi}{\lambda}\rho({t}_{N},\varphi_w)}]^T$, respectively. Denote by  $\boldsymbol{A}_w({\boldsymbol{r}},{\boldsymbol{t}}) = 	\mathbf{a}_r(\varphi_w,{\boldsymbol{r}})\mathbf{a}_t(\varphi_w,{\boldsymbol{t}})^H$ the response matrix for the $w$-th sensing target, and the received echo signal is given by
	\begin{equation}
		\boldsymbol{y}(m) = \sum_{w=1}^{W} \varepsilon_w\boldsymbol{A}_w({\boldsymbol{r}},{\boldsymbol{t}})\boldsymbol{x}(m)+\boldsymbol{n}_r(m),\forall w \in \mathcal{W},\label{echo}
	\end{equation}
	where $\varepsilon_w$  is the complex reflection coefficient for the $w$-th target, which captures both the round-trip path loss and radar cross section (RCS), and $\mathcal{W}\triangleq\{1,\dots,W\}$. $\boldsymbol{n}_r(m)\sim\mathcal{CN}(0,\sigma_r^2\boldsymbol{I}_N)$ is the AWGN at the BS. Both communication and dedicated radar waveforms can be exploited as probing signals since they are perfectly known by the BS. According to~\cite{zhang2024robust}, two groups of receiving filters should be designed to match the waveforms in~\eqref{hypotheses}. Denote by  $\boldsymbol{u}_{w,i}\in \mathbb{C}^{N\times 1}$  the receiving filter under $\mathcal{H}_i$, and the corresponding radar SINR for the $w$-th target can be calculated by
	\begin{align}
		&\Gamma_{w,i}(\boldsymbol{R}_X^i,\boldsymbol{t},\boldsymbol{r},\boldsymbol{u}_{w,i}) = \\\notag &\frac{|\alpha_w|^2\boldsymbol{u}_{w,i}^H\boldsymbol{A}_w(\boldsymbol{r},\boldsymbol{t})\boldsymbol{R}_X^i\boldsymbol{A}_w(\boldsymbol{r},\boldsymbol{t})^H\boldsymbol{u}_{w,i}}{\sum_{c\neq w}^{W}|\alpha_c|^2\boldsymbol{u}_{w,i}^H\boldsymbol{A}_c(\boldsymbol{r},\boldsymbol{t})\boldsymbol{R}_X^i\boldsymbol{A}_c(\boldsymbol{r},\boldsymbol{t})^H\boldsymbol{u}_{w,i}+\sigma_r^2\boldsymbol{u}_{w,i}^H\boldsymbol{u}_{w,i}},
	\end{align} 
	where  $\mathbb{E}\{|\varepsilon_w|^2\} = \alpha_w^2$. Note that in our design,
	we consider the target tracking stage, in which the target parameters including $\{\varphi_w\}_{w=1}^W$ and $\{|\alpha_w|^2\}_{w=1}^{W}$ have been roughly estimated in the previous stage. We assume that the target is quasi-static, so that the estimated parameters are sufficient for optimization design\cite{9652071}.
	\subsection{Detection Performance and Covertness Constraints}
	{Recall that} the sensing targets are assumed to be malicious Willies {who aim to distinguish between the two hypotheses $\mathcal{H}_0$ and $\mathcal{H}_1$}  based on received signals. In the following, we develop the optimal detection criteria and covertness constraints for both non-colluding and colluding Willies, respectively.
	\subsubsection{Non-colluding Willies}Under non-colluding scenarios, each Willie performs detection independently. Specifically, the signals received at Willie $w$ over $M$ time slots in one round are stacked into an observation vector, given by  $\boldsymbol{y}_w = [y_w(1), y_w(2), \dots, y_w(M)]^T \in \mathbb{C}^{M\times1}$, where each element $y_w(m)$ is expressed as
	\begin{equation}
		y_w(m) = \beta_w\mathbf{a}_t(\varphi_w,{\boldsymbol{t}})^H\boldsymbol{x}(m)+ n_w(m),\forall m\in\mathcal{M},\label{non-colluding_y}
	\end{equation}
	where $\beta_w$ denotes the corresponding path loss, and   $n_w(m)\sim\mathcal{CN}(0,\sigma_0^2),\forall w,$ is the AWGN at Willie $w$.   Denote by $\mathcal{D}_0$ and $\mathcal{D}_1$ the binary decisions in support of $\mathcal{H}_0$ and $\mathcal{H}_1$, respectively.  We assume that each Willie uses classical hypothesis testing with equal prior probabilities of each hypothesis being true~\cite{6584948}.  Then, the DEP at Willie $w$ can be given by $	\xi_w = \mathbb{P}(\mathcal{D}_0|\mathcal{H}_1) + \mathbb{P}(\mathcal{D}_1|\mathcal{H}_0),$ where $P_{MD} = \mathbb{P}(\mathcal{D}_0|\mathcal{H}_1)$ and $P_{FA} = \mathbb{P}(\mathcal{D}_1|\mathcal{H}_0)$ denote the miss detection probability (MDP) and the false alarm probability (FAP),  respectively. To achieve an optimal test that minimizes DEP $\xi_w$,  we assume that the likelihood  ratio test is performed  at each Willie~\cite{10090449}. Specifically, the optimal  test and corresponding DEP are derived as follows.
	
		\textit{Theorem 1: The optimal test for  Willie $w$ to detect the covert communication behaviours is derived as} 
	\begin{equation}
			\frac{\mathbb{P}(\boldsymbol{y}_w|\mathcal{H}_1)}{\mathbb{P}(\boldsymbol{y}_w|\mathcal{H}_0)} \overset{\mathcal{D}_1}{\underset{\mathcal{D}_0}{\gtrless}} 1 \Rightarrow ||\boldsymbol{y}_w||^2_2 \overset{\mathcal{D}_1}{\underset{\mathcal{D}_0}{\gtrless}} M\frac{\eta_{w,0} \eta_{w,1}}{\eta_{w,1} - \eta_{w,0}} \ln \frac{\eta_{w,1}}{\eta_{w,0}} \triangleq \varpi^*_w,
	\end{equation}
	\textit{where  $\mathbb{P}_{w,i} = \mathbb{P}(\boldsymbol{y}_w|\mathcal{H}_i) = \frac{1}{(\pi\eta_{w,i})^M}\exp(\frac{-||\boldsymbol{y}_w||^2_2}{\eta_{w,i}}), \forall i \in \{0,1\}$ is the probability density function (PDF) of $\boldsymbol{y}_w$ under hypothesis $\mathcal{H}_i$, $||\boldsymbol{y}_w||^2_2$ denotes the received signal power at Willie $w$,  $\eta_{w,0} = |\beta_w|^2\mathbf{a}_t(\varphi_w,{\boldsymbol{t}})^H\boldsymbol{R}_0\mathbf{a}_t(\varphi_w,{\boldsymbol{t}}) + \sigma_0^2$, and $\eta_{w,1} = \sum_{k=1}^{K}|\beta_w|^2  |\mathbf{a}_t(\varphi_w,{\boldsymbol{t}})^H\boldsymbol{w}_k|^2 +\eta_{w,0}$. Consequently, the minimum DEP  $\xi^\star_w$ is given by}
	\begin{subequations} \label{DEP_noncolluding}
		\begin{alignat}{2}
			\xi^\star_w &= \mathbb{P}(||\boldsymbol{y}_w||^2_2 \leq \varpi^*_w | \mathcal{H}_1) + \mathbb{P}(||\boldsymbol{y}_w||^2_2 \geq \varpi^*_w | \mathcal{H}_0) \\
			&= 1 - \underbrace{ \big(\frac{\gamma(M,\varpi^*_w/\eta_{w,0})}{\Gamma(M)} -\frac{\gamma(M,\varpi^*_w/\eta_{w,1})}{\Gamma(M)}\big)  }_{=~\mathcal{V}_T(\mathbb{P}_{w,0},\mathbb{P}_{w,1})} ,
		\end{alignat}
	\end{subequations}
	\textit{where \( \gamma (\cdot, \cdot) \) is the lower incomplete Gamma function given by \( \gamma (M, x) = \int_0^x e^{-t} t^{M-1} dt \), and $\Gamma(M) = (M-1)!$ is the Gamma function. $\mathcal{V}_T(\mathbb{P}_{w,0},\mathbb{P}_{w,1})$ is the total variation between PDFs $\mathbb{P}_{w,0}$ and $\mathbb{P}_{w,1}$.} 

	Based on the detection performance in~\eqref{DEP_noncolluding}, the covertness constraint for the $w$-th Willie can be given by 
	\begin{equation}
		\xi^\star_w \ge 1 - \epsilon_w,\forall w \in \mathcal{W},\label{CC1}
	\end{equation}
	where $\epsilon_w$ is the  given covertness level against Willie $w$~\cite{bash2013limits}.	Note that  a smaller $\epsilon_w$ corresponds to a more stringent covertness requirement for Willie $w$. Directly addressing the covertness constraint in~\eqref{CC1} is highly intractable  due to the lower incomplete Gamma functions. Fortunately, according to Pinsker's inequality in \cite{7447769}, we have $\mathcal{V}_T(\mathbb{P}_{w,0},\mathbb{P}_{w,1})\le \sqrt{\frac{1}{2}\mathcal{D}(\mathbb{P}_{w,0}||\mathbb{P}_{w,1})}$, where   $\mathcal{D}(\mathbb{P}_{w,0}||\mathbb{P}_{w,1})$ is the Kullback-Leibler (KL) divergence from $\mathbb{P}_{w,0}$ to $\mathbb{P}_{w,1}$, given by  
	\begin{equation}
		\mathcal{D}(\mathbb{P}_{w,0}||\mathbb{P}_{w,1}) = M\left(\ln\left(\frac{\eta_{w,1}}{\eta_{w,0}}\right) + \frac{\eta_{w,0}}{\eta_{w,1}} - 1\right).\label{KL}
	\end{equation}
	Therefore, the constraints in~\eqref{CC1} can be approximated as
	\begin{equation}\label{covert1}
		\mathcal{D}(\mathbb{P}_{w,0}||\mathbb{P}_{w,1}) \le 2\epsilon^2_w,\forall w \in \mathcal{W}.
	\end{equation}
	\subsubsection{Colluding Willies}In the colluding mode, each Willie delivers its observations to  a fusion center to jointly detect the presence of communication behaviours (e.g., multiple Willies are considered as a  powerful  {coordinated} node~\cite{11026809}). Different from~\eqref{non-colluding_y}, the received signal at colluding Willies in the $m$-th time slot can be expressed  as
	\begin{equation}\label{signal_warden_colluding}
		\boldsymbol{y}_F(m) = \boldsymbol{H}_F(\boldsymbol{t})\boldsymbol{x}(m)+\boldsymbol{n}_F(m),
	\end{equation}
	%		\begin{equation}\label{signal_warden_colluding}
		%			\boldsymbol{y}_F(m) = 
		%			\begin{cases} 
			%				\boldsymbol{H}_F(\boldsymbol{t})\boldsymbol{r}(m)+\boldsymbol{n}_F(m), \mathcal{H}_0 \\
			%				\boldsymbol{H}_F(\boldsymbol{t})(\sum_{k=1}^{K}\boldsymbol{w}_k\boldsymbol{s}_k(m)+\boldsymbol{r}(m))+\boldsymbol{n}_F(m), \mathcal{H}_1,
			%			\end{cases}
		%		\end{equation}
	where $\boldsymbol{H}_F(\boldsymbol{t}) = [\beta_1^*\mathbf{a}_t(\varphi_1,{\boldsymbol{t}}),\dots,\beta_W^*\mathbf{a}_t(\varphi_W,{\boldsymbol{t}})]^H\in\mathbb{C}^{W\times N}$, and $\boldsymbol{n}_F(m) = [n_1(m),\dots,n_W(m)]^T\in \mathbb{C}^{W\times1}$. The fusion center is utilized to make the detection decision by integrating received signals from all Willies. Let $\boldsymbol{Y}_F = [\boldsymbol{y}_F(1),\boldsymbol{y}_F(2),\dots,\boldsymbol{y}_F(M)]\in\mathbb{C}^{W\times M}$ denote the 
	received signal matrix in one round by Willies, the optimal test can be derived as follows.
	
	\textit{Theorem 2: The optimal test for colluding Willies to detect the covert communication behaviours is given by} 
	\begin{equation}\label{decision_colluding}
		||\boldsymbol{V}^{1/2}\boldsymbol{Y}_F||^2_F\overset{\mathcal{D}_1}{\underset{\mathcal{D}_0}{\gtrless}} M\ln \dfrac{\text{det}(\boldsymbol{\Lambda}_1)}{\text{det}(\boldsymbol{\Lambda}_0)} \triangleq \chi,
	\end{equation}
	\textit{where $\boldsymbol{\Lambda}_0 = \boldsymbol{H}_F(\boldsymbol{t})\boldsymbol{R}_0\boldsymbol{H}_F(\boldsymbol{t})^H+\sigma_0^2\boldsymbol{I}_W\in\mathbb{C}^{W\times W}$, $\boldsymbol{\Lambda}_1 = \boldsymbol{H}_F(\boldsymbol{t})\boldsymbol{W}\boldsymbol{W}^H\boldsymbol{H}_F(\boldsymbol{t})^H+\boldsymbol{\Lambda}_0\in\mathbb{C}^{W\times W}$,  \textit{and} $\boldsymbol{V} = \boldsymbol{\Lambda}_0^{-1} - \boldsymbol{\Lambda}_1^{-1}\in\mathbb{C}^{W\times W}$}. \textit{The corresponding minimum detection error probability ${\xi}_F^\star$ is given by}
	\begin{equation}\label{colludingMDP}
		{\xi}_F^\star = 1 - \big(\underbrace{F(\chi|\frac{1}{\lambda_1^0},\dots,\frac{1}{\lambda_W^0})-F(\chi|\frac{1}{\lambda_1^1},\dots,\frac{1}{\lambda_W^1})}_{=~\mathcal{V}_T(\mathbb{P}_{F,0},\mathbb{P}_{F,1})}\big) ,
	\end{equation}
	\textit{where $\lambda_1^0,\dots,\lambda_W^0$ and ${\lambda}_1^1,\dots,{\lambda}_W^1$ are the eigenvalues of $\boldsymbol{\Lambda}_0^{1/2}\boldsymbol{V}\boldsymbol{\Lambda}_0^{1/2}$ and $\boldsymbol{\Lambda}_1^{1/2}\boldsymbol{V}\boldsymbol{\Lambda}_1^{1/2}$,  respectively.  $F(\chi|\frac{1}{{\lambda}_1^i},\dots,\frac{1}{{\lambda}_W^i})$ is defined in~\eqref{PDF_colluding}, shown at the top of next page. Besides, $\mathcal{V}_T(\mathbb{P}_{F,0},\mathbb{P}_{F,1})$ is the total variation between PDFs of $\boldsymbol{Y}_F$ under $\mathcal{H}_0$ and $\mathcal{H}_1$, where  $\mathbb{P}_{F,i},\forall i \in \{0,1\}$ is elaborated in~\eqref{PDF_YF}.}
	\begin{figure*}[tp]	 	
		\begin{equation} \label{PDF_colluding}
			F\left ({{x|{\frac{1}{\lambda _{1}^i}}, \ldots, \frac{1}{{\lambda _{W}^i}}}}\right ) = 1 - \left ({{\prod \limits _{j = 1}^{W} \frac{1}{(\lambda _{j}^i)^{M}} }}\right )\sum \limits _{k = 1}^{W} {\sum \limits _{l = 1}^{M} {\left ({{\frac {\partial ^{l - 1}}{\partial {{\left ({{ - {\frac{1}{\lambda _{k}^i}}}}\right )}^{l - 1}}}\left ({{\prod \limits _{j = 1,j \ne k}^{W} {\frac {{\left ({{\frac{1}{{\lambda _{j}^i}} - \frac{1} {\lambda _{k}^i}}}\right )}^{ - M}}{\frac{1}{\lambda _{k}^i}}} }}\right )}}\right )\frac {{x^{M - l}}{e^{ - {\frac{1}{\lambda _{k}^i}}x}}}{\left ({{M - l}}\right )!(l - 1)!}} },\forall i.
		\end{equation}
		\hrule
	\end{figure*}
	
	\textit{Proof: Please refer to Appendix A.}\hfill$\Box$
	
	Similar to~\eqref{CC1}, the covertness constraint for colluding Willies can be given by
	\begin{equation}
		{\xi}_F^\star \ge 1 - \epsilon_F,\label{CC2}
	\end{equation}
	where $\epsilon_F$ is the covertness level against colluding detection~\cite{bash2013limits}. 
	The expression of the theoretical minimum DEP in~\eqref{colludingMDP} is rather intractable to guide further system design.  Similar to the non-colluding case,  $\mathcal{V}_T(\mathbb{P}_{F,0},\mathbb{P}_{F,1})$ is upper bounded by $\mathcal{V}_T(\mathbb{P}_{F,0},\mathbb{P}_{F,1})\le \sqrt{\frac{1}{2}\mathcal{D}(\mathbb{P}_{F,0}||\mathbb{P}_{F,1})}$, where  the KL divergence  $\mathcal{D}(\mathbb{P}_{F,0}||\mathbb{P}_{F,1})$ is derived as 
	\begin{equation}
		\mathcal{D}(\mathbb{P}_{F,0}||\mathbb{P}_{F,1}) = M\left(\text{Tr}(\boldsymbol{\Lambda}_1^{-1}\boldsymbol{\Lambda}_0)-\ln\det(\boldsymbol{\Lambda}_1^{-1}\boldsymbol{\Lambda}_0)-W\right).
	\end{equation}
	
	As such, the constraint in~\eqref{CC2} can be approximated as 
	\begin{equation}\label{covert2}
		\mathcal{D}(\mathbb{P}_{F,0}||\mathbb{P}_{F,1}) \le 2\epsilon_F^2.
	\end{equation}
	
\subsection{Problem Formulation}	
	In this paper, we aim to maximize the covert sum communication rate by jointly designing the transmit beamforming vectors, receiving filter, and transceiver antenna placement. The optimization problem can be formulated as  
	\begin{subequations}\label{Problem1}
		\begin{alignat}{2}
			&\underset{ \boldsymbol{W}, \boldsymbol{R}_0,  {\boldsymbol{r}},{\boldsymbol{t},\{\boldsymbol{u}_{w,i}\}}}{\max} \quad \sum_{k=1}^{K}\log_2(1+\gamma_k) \label{P1_rate}\\
			&\,\text {s.t.}~ \Gamma_{w,i}(\boldsymbol{R}_X^i,\boldsymbol{t},\boldsymbol{r},\boldsymbol{u}_{w,i}) \ge \Gamma_w,\forall w \in \mathcal{W},i\in\{0,1\},\label{P1_SNR}\\
			&\hphantom {s.t.~} t_1\ge 0, t_{N}\le D, r_1\ge 0,r_{N}\le D,\label{P1_region}\\
			&\hphantom {s.t.~}  t_n - t_{n-1}\ge d, r_n-r_{n-1}\ge d, 2\le n \le N, \label{P1_distance}\\
			&\hphantom {s.t.~}
			\sum_{k=1}^{K}\boldsymbol{w}_k^H\boldsymbol{w}_k + \text{Tr}(\boldsymbol{R}_0) \le P_t,\label{P1_Power}\\
			&\hphantom {s.t.~}
			\boldsymbol{R}_0\succeq 0,\label{P1_R0}\\
			&\hphantom {s.t.~} \eqref{covert1}~\text{or}~\eqref{covert2}, \label{P1_covert}
		\end{alignat} 
	\end{subequations} 
	where $\Gamma_w$ is the radar SINR  threshold for the $w$-th target, $d$ represents the minimum distance between MAs to prevent coupling effects, $P_t$ is the maximum transmission power, and the constraint in~\eqref{P1_covert} guarantee the covertness level of confidential transmission under non-colluding or colluding Willie scenarios. We note that the problem in~\eqref{Problem1} is computationally intractable due to the highly non-convex objective function and tight coupling among optimization variables. The dependence of KL divergence on beamforming vectors and antenna placement across different detection modes further exacerbates the difficulty of algorithm development. In the following sections, we first propose a BCD-based algorithm for non-colluding Willies, and then modify it to address the collusion detection problem.

\vspace{-2mm}
	\section{Proposed BCD Algorithm}
 In this section, we address the non-colluding detection problem with constraints in~\eqref{covert1}. Specifically, we first reformulate the objective function in~\eqref{P1_rate} into a more tractable form by using the Lagrangian dual transformation technique. Then, we develop a BCD-based algorithm, the details of which are elaborated as follows. 
	
	\vspace{-4mm}
	\subsection{Problem Reformulation}
	Based on the Lagrangian dual transformation method in~\cite{10772590}, we equivalently transform the original objective function in~\eqref{P1_rate} as
	\begin{align}
		\mathcal{F}_1(\boldsymbol{W},\boldsymbol{R}_0,\boldsymbol{t},\boldsymbol{\rho})& =  \sum_{k=1}^{K}\{\ln (1+\rho_k)-\rho_k +\\\notag
		&\frac{(1+\rho_k)|\boldsymbol{h}_k^H(\boldsymbol{t})\boldsymbol{w}_k|^2}{\sum_{i=1}^{K}|\boldsymbol{h}_k^H(\boldsymbol{t})\boldsymbol{w}_i|^2+\boldsymbol{h}_k^H(\boldsymbol{t})\boldsymbol{R}_0\boldsymbol{h}_k(\boldsymbol{t})+\sigma_k^2}\},
	\end{align} %$\mathcal{F}_1(\boldsymbol{W},\boldsymbol{R}_0,\boldsymbol{t},\boldsymbol{\rho}) =  \sum_{k=1}^{K}\{\ln (1+\rho_k)-\rho_k +
%	\frac{(1+\rho_k)|\boldsymbol{h}_k^H(\boldsymbol{t})\boldsymbol{w}_k|^2}{\sum_{i=1}^{K}|\boldsymbol{h}_k^H(\boldsymbol{t})\boldsymbol{w}_i|^2+\boldsymbol{h}_k^H(\boldsymbol{t})\boldsymbol{R}_0\boldsymbol{h}_k(\boldsymbol{t})+\sigma_k^2}\},$
	%	\begin{align}
		%		\mathcal{F}_1(\boldsymbol{W},&\boldsymbol{R}_0,\boldsymbol{t},\boldsymbol{\rho}) =  \sum_{k=1}^{K}\{\ln (1+\rho_k)-\rho_k +\notag \\
		%		&\frac{(1+\rho_k)|\boldsymbol{h}_k^H(\boldsymbol{t})\boldsymbol{w}_k|^2}{\sum_{i=1}^{K}|\boldsymbol{h}_k^H(\boldsymbol{t})\boldsymbol{w}_i|^2+\boldsymbol{h}_k^H(\boldsymbol{t})\boldsymbol{R}_0\boldsymbol{h}_k(\boldsymbol{t})+\sigma_k^2}\},\label{OBJ1}
		%	\end{align}
	where $\boldsymbol{\rho} = [\rho_1,\rho_2,\dots,\rho_K]^T\in \mathbb{R}^{K\times1}$ is a slack variable vector. It can be readily seen that the reformulated objective function is convex with respect to (w.r.t.) $\rho_k$, and thus the optimal $\rho_k^\star$ can be derived by checking the first-order optimality  condition, i.e.,
	\begin{equation}
		\rho_k^\star = \gamma_k,\forall k.\label{rho_update}
	\end{equation}
	 Next, we note that with  $\boldsymbol{\rho}$ being fixed, only the last term of $\mathcal{F}_1(\boldsymbol{W},\boldsymbol{R}_0,\boldsymbol{t},\boldsymbol{\rho})$, which is in a sum-of-ratio form, is involved in the optimization on $\boldsymbol{W},\boldsymbol{R}_0$ and $\boldsymbol{t}$. To deal with this issue, we first introduce an auxiliary variable $\boldsymbol{\upsilon} \triangleq [\upsilon_1,\dots,\upsilon_K]^T \in \mathbb{C}^{K\times1}$. Then, the quadratic transformation method is employed to recast $\mathcal{F}_1(\boldsymbol{W},\boldsymbol{R}_0,\boldsymbol{t},\boldsymbol{\rho})$ as 
	\begin{align}
		\mathcal{F}_2(&\boldsymbol{W},\boldsymbol{R}_0,\boldsymbol{t},\boldsymbol{\upsilon}) =  \sum_{k=1}^{K}\{2(1+\rho_k)\upsilon_k\sqrt{|\boldsymbol{h}_k^H({\boldsymbol{t}})\boldsymbol{w}_k|^2}\notag\\
		&-(1+\rho_k)|\upsilon_k|^2\left(\boldsymbol{h}_k^H(\boldsymbol{t})\boldsymbol{R}_X^1\boldsymbol{h}_k(\boldsymbol{t})+\sigma_k^2\right)\} + \text{const},\label{F2}
	\end{align}
	where $\text{const}$ refers to a constant term that is independent of optimization variables.  However, $	\mathcal{F}_2(\boldsymbol{W},\boldsymbol{R}_0,\boldsymbol{t},\boldsymbol{\upsilon})$ is still non-concave due to the coupling of optimization variables. Therefore, we  propose a BCD-based 
	algorithm to obtain an efficient solution.
	\vspace{-2mm}
	\subsection{Updating Auxiliary Variable}
	With $\boldsymbol{W},\boldsymbol{R}_0,\boldsymbol{t},\boldsymbol{r}$, and $\{\boldsymbol{u}_{w,i}\}$  being fixed, it is observed that  $\mathcal{F}_2(\boldsymbol{W},\boldsymbol{R}_0,\boldsymbol{t},\boldsymbol{\upsilon})$ is convex w.r.t. $\upsilon_k,\forall k$, and the closed-form solution can be given by 
	\begin{equation}
		\upsilon_k^\star = \frac{\sqrt{|\boldsymbol{h}_k^H(\boldsymbol{t})\boldsymbol{w}_k|^2}}{\boldsymbol{h}_k^H(\boldsymbol{t})\boldsymbol{R}_X^1\boldsymbol{h}_k(\boldsymbol{t})+\sigma_k^2},\forall  k.\label{v}
	\end{equation}
	\subsection{Updating Transmit Beamforming}
	With all other variables being fixed, we focus on optimization on transmit beamforming $\boldsymbol{W}$ and $\boldsymbol{R}_0$. Regarding the constraints in \eqref{covert1}, since $f(x) = \ln x + \frac{1}{x} - 1$ is monotonically increasing for $x \in [1, +\infty)$ and given that $\frac{\eta_{w,1}}{\eta_{w,0}} \ge 1$ always holds, the covertness constraints can be equivalently recast as 
	\begin{equation}\label{covert_noncolluding}
		\frac{\eta_{w,1}}{\eta_{w,0}} \le \kappa_w,\forall w\in\mathcal{W},
	\end{equation} 
	where $\kappa_w$ is the unique numerical solution of the equation $\mathcal{D}(\mathbb{P}_{w,0}||\mathbb{P}_{w,1}) =  2\epsilon^2_w$ in the interval $[1,+\infty)$. 
	Then, it is observed that the non-convexity of the problem in~\eqref{Problem1} lies in the quadratic terms w.r.t. $\{\boldsymbol{w}_k\}_{k=1}^K$ in~\eqref{P1_SNR},~\eqref{F2}, and~\eqref{covert_noncolluding}.  To deal with this issue, the SDR method can be  employed~\cite{5447068}. Specifically, we first construct auxiliary optimization variables  $\{\boldsymbol{R}_k\}_{k=1}^K$ with $\boldsymbol{R}_k = \boldsymbol{w}_k\boldsymbol{w}_k^H$, which is a rank-one positive semidefinite matrix. Combining~\eqref{F2}, the problem in~\eqref{Problem1} can be recast as
\begin{small}
		\begin{subequations}\label{Problem2}
			\begin{alignat}{2}
				&\underset{ \{\boldsymbol{R}_k\}_{k=0}^{K}}{\max} ~\sum_{k=1}^{K}\Big\{2(1+\rho_k)\upsilon_k\sqrt{\boldsymbol{h}_k^H({\boldsymbol{t}}){\boldsymbol{R}}_k\boldsymbol{h}_k({\boldsymbol{t}})}\notag\\
				&\quad\quad\quad-(1+\rho_k)|\upsilon_k|^2\left(\boldsymbol{h}_k^H(\boldsymbol{t})\boldsymbol{R}_X^1\boldsymbol{h}_k(\boldsymbol{t})+\sigma_k^2\right)\Big\} \label{P2_OBJ}\\
				&\,\text {s.t.}~
				|\alpha_w|^2\text{Tr}\Big(\boldsymbol{R}_X^i\boldsymbol{\Xi}_{w,w}^i\Big)-\sigma_r^2\Gamma_w\boldsymbol{u}_{w,i}^H\boldsymbol{u}_{w,i}\ge\notag\\
				&~~~~~~~~~~\Gamma_w\sum_{c\neq w}^{W}|\alpha_c|^2\text{Tr}\Big(\boldsymbol{R}_X^i\boldsymbol{\Xi}_{w,c}^i\Big),\forall w\in\mathcal{W},\forall i\in\{0,1\},\label{P2_radar_SINR}\\
				&\hphantom {s.t.~}|\beta_w|^2\text{Tr}\Big(\boldsymbol{R}_X^1\boldsymbol{\Upsilon}_w\Big) + (1-\kappa_w)\sigma_w^2\le\kappa_w|\beta_w|^2\text{Tr}\Big(\boldsymbol{R}_0\boldsymbol{\Upsilon}_w\Big),\label{P2_covert}\\
				&\hphantom {s.t.~}\boldsymbol{R}_0\succeq0,~\sum_{k=0}^{K}\text{Tr}\Big(\boldsymbol{R}_k\Big)\le P_t,\label{P2_power} \\
				&\hphantom {s.t.~} \boldsymbol{R}_k \succeq 0,~\text{rank}(\boldsymbol{R}_k) = 1,1\le k \le K, \label{P2_rank}
			\end{alignat} 
		\end{subequations} 
	\end{small}where $\boldsymbol{\Xi}_{w,c}^i = \boldsymbol{A}_c(\boldsymbol{r},\boldsymbol{t})^H\boldsymbol{u}_{w,i}\boldsymbol{u}_{w,i}^H\boldsymbol{A}_c(\boldsymbol{r},\boldsymbol{t}),\boldsymbol{\Upsilon}_w =\mathbf{a}_t(\varphi_w,\boldsymbol{t}) \mathbf{a}_t^H(\varphi_w,\boldsymbol{t}),\forall w,c \in\mathcal{W},\forall i \in \{0,1\}$. By dropping the rank-one constraints in~\eqref{P2_rank}, the problem in~\eqref{Problem2} is a semidefinite program (SDP) and can be solved by using the CVX tool\cite{grant2008cvx}. Denote by  $\{\tilde{\boldsymbol{R}}_k\}_{k=0}^K$ the optimal solutions for the relaxed problem. Here, we would like to note that if $\{\tilde{\boldsymbol{R}}_k\}_{k=1}^K$ is exactly rank-one, the solution to the relaxed problem is also an optimal solution to the original non-convex problem. While such relaxations are not necessarily tight,  we can always construct  a  closed-form rank-one  solution based on  $\{\tilde{\boldsymbol{R}}_k\}_{k=0}^K$ in a heuristic manner. Specifically, we can  obtain $\bar{\boldsymbol{R}}_0$ and rank-one $\{{\bar{\boldsymbol{R}}}_k\}_{k=1}^K$ to optimally enhance the covert rate under $\mathcal{H}_1$ via
		\begin{align}
			&\boldsymbol{R}_X^1 = \sum_{k=0}^{K}\tilde{\boldsymbol{R}}_k,~{\boldsymbol{w}}_k = (\boldsymbol{h}_k^H(\boldsymbol{t})\tilde{\boldsymbol{R}}_k\boldsymbol{h}_k(\boldsymbol{t}))^{-1/2}\tilde{\boldsymbol{R}}_k\boldsymbol{h}_k(\boldsymbol{t}), \notag\\
			&{\bar{\boldsymbol{R}}}_k = {\boldsymbol{w}}_k{\boldsymbol{w}}_k^H,~ \bar{\boldsymbol{R}}_0 = \boldsymbol{R}_X^1 - \sum_{k=1}^{K}\bar{\boldsymbol{R}}_k,1 \le k\le K,\label{construct}
		\end{align}
		
	 Proof: Please refer to Appendix B.
		\subsection{Updating Transmit Antenna Placement}
		In this subsection, we carry out optimization on $\boldsymbol{t}$. It is worth noting that the non-convexity lies in  the objective function $\mathcal{F}_2(\boldsymbol{t})$ in~\eqref{F2}, and the constraints in~\eqref{P1_SNR} and~\eqref{covert_noncolluding}. To handle this challenge, we introduce a projected gradient descent algorithm, with Nesterov’s acceleration strategy being  incorporated to speed up the convergence\cite{10772590}. Let $\nabla{\mathcal { F}_2}(\boldsymbol{t})\in\mathbb{C}^{N\times 1}$ be the gradient vector at $\boldsymbol{t}$,  and  the antenna position $\boldsymbol{t}$ is updated by the following steps
		\begin{subequations} 
			\begin{alignat}{2}
				&\text{(Step. 1)}~\boldsymbol{m}^{l+1} ={{ {\boldsymbol z}^{l} + \eta^l \nabla  {\mathcal { F}_2}({\boldsymbol z}^{l}) }}, \label{a}\\
				&\text{(Step. 2)}~\boldsymbol{t}^{l+1}  = \text{arg}~ \underset{\boldsymbol{t}}{\min}~ ||\boldsymbol{t}-\boldsymbol{m}^{l+1}||_2^2\notag \\ &~~~~~~~~~~~~~~~~~~~~\text{s.t.}~\eqref{P1_SNR},\eqref{P1_region},\eqref{P1_distance},\eqref{covert_noncolluding}\label{b},\\
				&\text{(Step. 3)}~{\boldsymbol z}^{l+1}  = {\boldsymbol t}^{l+1} + \zeta _{l+1}({\boldsymbol t}^{l+1} - {\boldsymbol t}^{l}),\label{c}
			\end{alignat}
		\end{subequations}
		where $\boldsymbol{m}^{l+1}\in\mathbb{C}^{N\times1}$ is an auxiliary variable, and $\eta^l\ge0$ is the descent step length, which can be calculated by the backtracking line search method. {The superscript $l$ indicates the iteration index.} Here, $\nabla  {\mathcal { F}_2}({\boldsymbol z}^{l})$ is the gradient of $\mathcal{F}_2(\boldsymbol{t})$ at $\boldsymbol{z}^l$,  $\zeta _{l+1} = \frac{\alpha_{l+1}-1}{\alpha_{l+1}}$, and $\alpha_{l+1} = \frac{1+\sqrt{1+4\alpha_l^2}}{2}$ with $\alpha_1 = 0.1.$ For  Step. 1 in~\eqref{a}, we define 
		that $\tilde{\mathcal{F}}_{i,j}(\boldsymbol{t}) \triangleq \boldsymbol{h}_i^H(\boldsymbol{t})\boldsymbol{R}_j\boldsymbol{h}_i(\boldsymbol{t})$, where $1\le i\le K,0\le j\le K$. Thus, the gradient vector $\nabla\mathcal{F}_2(\boldsymbol{t})$ can be given by 
			\begin{align}
				\nabla\mathcal{F}_2(\boldsymbol{t}) = \sum_{k=1}^{K}&(1+\rho_k)v_k\frac{1}{\sqrt{\tilde{\mathcal{F}}_{k,k}(\boldsymbol{t})}}\nabla \tilde{\mathcal{F}}_{k,k}(\boldsymbol{t})\notag \\
				&-\sum_{k=1}^{K}\sum_{j=0}^{K}(1+\rho_k)|v_k|^2\nabla \tilde{\mathcal{F}}_{k,j}(\boldsymbol{t}),\label{gradient}
			\end{align}
where $\nabla 	\tilde{\mathcal{F}}_{i,j}(\boldsymbol{t}) \in \mathbb{C}^{N\times 1}$ denotes the gradient vector of $\tilde{\mathcal{F}}_{i,j}(\boldsymbol{t}) $  at $\boldsymbol{t}$. Please refer to  Appendix C for derivation of  $\nabla 	\tilde{\mathcal{F}}_{i,j}(\boldsymbol{t})$. Then, we move on to deal with the problem in Step. 2. The problem is intractable due to the constraints in~\eqref{P1_SNR} and~\eqref{covert_noncolluding}. To deal with these issues, the SCA method can be employed. According to the second-order Taylor expansion theorem in~\cite{boyd2004convex}, the non-convex parts of constraints in~\eqref{P1_SNR} and~\eqref{covert_noncolluding} can be respectively approximated as
	\begin{subequations} \label{P4_approximation}
	\begin{alignat}{2}
		\tilde{\Gamma}_{w,i}(\boldsymbol{t})\geq~&\tilde{\Gamma}_{w,i}(\boldsymbol{t}^l) + \nabla\tilde{\Gamma}_{w,i}(\boldsymbol{t}^l)^T(\boldsymbol{t} - \boldsymbol{t}^l)-\frac{\tilde{\delta}_{w,i}}{2}||{\boldsymbol{t}}-{\boldsymbol{t}}^{l}||_2^2,\label{P4_SNR_app}\\
		\mathcal{G}_w(\boldsymbol{t})\leq&\mathcal{G}_w(\boldsymbol{t}^l) + \nabla \mathcal{G}_w(\boldsymbol{t}^l)^T (\boldsymbol{t}-\boldsymbol{t}^l)+\frac{\bar{\delta}_{w}}{2}||\boldsymbol{t}-{\boldsymbol{t}}^{l}||_2^2,\label{P4_covert_app}
	\end{alignat}
\end{subequations}
%			\begin{align}
%				\tilde{\Gamma}_{w,i}(\boldsymbol{t})\geq~&\tilde{\Gamma}_{w,i}(\boldsymbol{t}^l) + \nabla\tilde{\Gamma}_{w,i}(\boldsymbol{t}^l)(\boldsymbol{t} - \boldsymbol{t}^l)-\frac{\tilde{\delta}_{w,i}}{2}||{\boldsymbol{t}}-{\boldsymbol{t}}^{l}||,\label{P4_SNR_app}\\
%				\mathcal{G}_w(\boldsymbol{t})\leq&\mathcal{G}_w(\boldsymbol{t}^l) + \nabla \mathcal{G}_w(\boldsymbol{t}^l) (\boldsymbol{t}-\boldsymbol{t}^l)+\frac{\bar{\delta}_{w}}{2}||\boldsymbol{t}-{\boldsymbol{t}}^{l}||_2^2,\label{P4_covert_app}
%			\end{align}
where $\boldsymbol{t}^l$ is the obtained APV in the $l$-th iteration, $\tilde{\Gamma}_{w,i}(\boldsymbol{t}) = \boldsymbol{u}_{w,i}^H(|\alpha_w|^2\boldsymbol{A}_w(\boldsymbol{r},\boldsymbol{t})\boldsymbol{R}_X^i\boldsymbol{A}_w(\boldsymbol{r},\boldsymbol{t})^H-\sum_{c\neq w}^{W}\Gamma_w|\alpha_c|^2\boldsymbol{A}_c(\boldsymbol{r},\boldsymbol{t})
\boldsymbol{R}_X^i\boldsymbol{A}_c(\boldsymbol{r},\boldsymbol{t})^H-\sigma_r^2\Gamma_{w}\boldsymbol{I}_N)\boldsymbol{u}_{w,i}$, and  $\mathcal{G}_w(\boldsymbol{t}) = \boldsymbol{a}_t^H(\varphi_w,\boldsymbol{t})\left(|\beta_w|^2\boldsymbol{R}_X^1-\kappa_w|\beta_w|^2\boldsymbol{R}_{X}^0\right)\boldsymbol{a}_t(\varphi_w,\boldsymbol{t})+(1-\kappa_w)\sigma_0^2.$ Note that the positive real numbers $\tilde{\delta}_{w,i}$ and  $\bar{\delta}_{w}$ are selected to satisfy $\tilde{\delta}_{w,i}\boldsymbol{I}_N\succeq\nabla^2\tilde{\Gamma}_{w,i}(\boldsymbol{t})$ and $\bar{\delta}_{w}\boldsymbol{I}_N\succeq\nabla^2\mathcal{G}_w(\boldsymbol{t})$, with $\nabla^2\tilde{\Gamma}_{w,i}(\boldsymbol{t})\in\mathbb{C}^{N\times N}$ and $\nabla^2\mathcal{G}_w(\boldsymbol{t})\in\mathbb{C}^{N\times N}$ being the Hessian matrices, respectively.  Please refer to  Appendix D for the construction of $\nabla\tilde{\Gamma}_{w,i}(\boldsymbol{t}^l)$, $\nabla \mathcal{G}_w(\boldsymbol{t}^l)$, $\tilde{\delta}_{w,i}$ and  $\bar{\delta}_{w}$. Thus, combing~\eqref{P4_SNR_app} and~\eqref{P4_covert_app}, the problem in~\eqref{b} can be approximated  as follows
		\begin{small}
			\begin{subequations}\label{Problem5}
				\begin{alignat}{2}
					&\underset{{\boldsymbol{t}}}{\min} \quad ||\boldsymbol{t}-\boldsymbol{m}^{l+1}||_2^2 \\
					&\,\text {s.t.}~t_1 \ge 0, t_N \le D,t_n - t_{n-1}\ge d, 2 \le n \le N,\label{P5_distance}\\
					&\hphantom {s.t.~}   	\tilde{\Gamma}_{w,i}(\boldsymbol{t}^l) + \nabla\tilde{\Gamma}_{w,i}(\boldsymbol{t}^l)^T(\boldsymbol{t} - \boldsymbol{t}^l)-\frac{\tilde{\delta}_{w,i}}{2}||{\boldsymbol{t}}-{\boldsymbol{t}}^{l}||_2^2\ge 0,\notag\\
					&\hphantom {s.t.~}~~~~~~~~~~~~~~~~~~~~~~~~~~~~~~~~~~~~~~~   \forall w \in \mathcal{W},\forall i \in \{0,1\},\label{P5_SINR}\\
					&\hphantom {s.t.~}~ \mathcal{G}_w(\boldsymbol{t}^l) + \nabla \mathcal{G}_w(\boldsymbol{t}^l)^T (\boldsymbol{t}-\boldsymbol{t}^l)+\frac{\bar{\delta}_{w}}{2}||\boldsymbol{t}-{\boldsymbol{t}}^{l}||_2^2\le 0,\forall w\in \mathcal{W},\label{P5_covert}
				\end{alignat} 
			\end{subequations} 
		\end{small}
		which is convex and can be solved by using the CVX tool.
		\vspace{-2mm}
		\subsection{Updating Receive Filter and Antenna Placement}
		Note that the objective function in~\eqref{F2} is independent of $\boldsymbol{r}$ and $\{\boldsymbol{u}_{w,i}\}$, which indicates that the receiver design is a feasibility-check problem and the solution will not directly affect $	\mathcal{F}_2(\boldsymbol{W},\boldsymbol{R}_0,\boldsymbol{t},\boldsymbol{\upsilon})$. To provide additional degrees of freedom (DoFs) for optimization on other variables, we propose to maximize the minimum radar SINR for the receiver design. By introducing an auxiliary variable $\mu$, the problem in~\eqref{Problem1} can be recast as 
		\begin{subequations}\label{Problem6}
			\begin{alignat}{2}
				&\underset{{\boldsymbol{r}},\{\boldsymbol{u}_{w,i}\},\mu}{\max}~\mu  \\
				&\,\text {s.t.}~~\Gamma_{w,i}(\boldsymbol{r},\boldsymbol{u}_{w,i})\ge \mu,\forall w\in \mathcal{W},\forall i \in \{0,1\},\label{P6_SINR}\\
				&\hphantom {s.t.~}  r_1\ge 0,r_N\le D, r_n - r_{n-1}\ge d, 2\le n \le N.
			\end{alignat} 
		\end{subequations} 
		As $\boldsymbol{r}$ and $\boldsymbol{u}_{w,i}$ are coupled, we likewise employ the BCD algorithm to address this problem. However, with $\{\boldsymbol{u}_{w,i}\}$ being fixed, the problem~\eqref{Problem6} w.r.t. $\boldsymbol{r}$ is still intractable due to the non-convexity of the constraints in~\eqref{P6_SINR}. To deal with this challenge, we employ the Dinkelbach  transformation in~\cite{dinkelbach1967nonlinear} to rewrite the constraints in~\eqref{P6_SINR} into a more favourable polynomial expression as
		\begin{equation}
			\bar{\Gamma}_{w,i}(\boldsymbol{r})\ge \mu,\forall w\in \mathcal{W},\forall i \in\{0,1\},\label{Dinkelbach}
		\end{equation}
		where $\bar{\Gamma}_{w,i}(\boldsymbol{r}) = |\alpha_w|^2\boldsymbol{u}_{w,i}^H\boldsymbol{A}_w(\boldsymbol{r},\boldsymbol{t})\boldsymbol{R}_X^i\boldsymbol{A}_w(\boldsymbol{r},\boldsymbol{t})^H\boldsymbol{u}_{w,i}-\varrho(\sum_{c\neq w}^{W}|\alpha_c|^2\boldsymbol{u}_{w,i}^H\boldsymbol{A}_c(\boldsymbol{r},\boldsymbol{t})\boldsymbol{R}_X^i\boldsymbol{A}_c(\boldsymbol{r},\boldsymbol{t})^H\boldsymbol{u}_{w,i}+\sigma_r^2\boldsymbol{u}_{w,i}^H\boldsymbol{u}_{w,i})$, and $\varrho \in\mathbb{R}$ is the auxiliary variable. Since the constraints in~\eqref{Dinkelbach} is still non-convex w.r.t. $\boldsymbol{r}$, we follow the idea of second-order Taylor expansion theorem to approximate the constraints in~\eqref{Dinkelbach} as 
		\begin{equation}
			\bar{\Gamma}_{w,i}(\boldsymbol{r}^l)  + \nabla\bar{\Gamma}_{w,i}(\boldsymbol{r}^l)^T(\boldsymbol{r} - \boldsymbol{r}^l)-\frac{\bar{\delta}_{w,i}}{2}||{\boldsymbol{r}}-{\boldsymbol{r}}^{l}||_2^2\ge \mu,
		\end{equation}
	which is convex, and can be solved by using the CVX tool. Then,  $\varrho$ can be updated by
	\begin{equation}
		\varrho' = \underset{w,i}{\min}~ {\Gamma}_{w,i}(\boldsymbol{r},\boldsymbol{u}_{w,i}).\label{varrho_update}
	\end{equation}The procedures for optimization on $\boldsymbol{r}$ are summarized in \textit{Algorithm 1}.
	
	 As for $\{\boldsymbol{u}_{w,i}\}$, it can be readily observed that the corresponding subproblem for each filter is a generalized Rayleigh quotient maximization problem~\cite{2025arXiv251009949Y}. The optimal solution $\boldsymbol{u}_{w,i}^\star$ can be directly given by the eigenvector associated with the largest eigenvalue of the matrix $\left(\sum_{c\neq w}^{W}|\alpha_c|^2\boldsymbol{A}_c(\boldsymbol{r},\boldsymbol{t})\boldsymbol{R}_X^i\boldsymbol{A}_c(\boldsymbol{r},\boldsymbol{t})^H+\sigma_r^2\boldsymbol{I}_N\right)^{-1}|\alpha_w|^2\\\boldsymbol{A}_w(\boldsymbol{r},\boldsymbol{t})\boldsymbol{R}_X^i\boldsymbol{A}_w(\boldsymbol{r},\boldsymbol{t})^H$. The overall BCD-based optimization algorithm for non-colluding Willies is summarized in \textit{Algorithm 2}. The convergence behaviour and complexity analysis are presented in Section IV.
	 \begin{small}
	 		\begin{algorithm}[t]
	 		\caption{Dinkelbach Transformation-based Algorithm for Problem in~\eqref{Problem6}}
	 		\begin{algorithmic}[1] 
	 			\STATE \textbf{Initialize}: $\boldsymbol{r}^{\iota},{\rho}^{\iota}$, and set $\iota=0$.
	 			\REPEAT		 
	 			\STATE Update $\boldsymbol{r}^{\iota+1}$ and $\mu^{\iota+1}$ by solving Problem in~\eqref{Problem6};
	 			\STATE Update $\varrho^{\iota+1}$  via~\eqref{varrho_update};
	 			\STATE Let $\iota = \iota+1$;
	 			\UNTIL Exist conditions are met.
	 			\RETURN  $\boldsymbol{r}^\star$.
	 		\end{algorithmic} 
	 	\end{algorithm}
	 \end{small}

		\begin{small}
			\begin{algorithm}[t]
				\caption{The overall BCD Algorithm for Problem in~\eqref{Problem1}}
				\begin{algorithmic}[1] 
					\STATE \textbf{Initialize}: $\boldsymbol{W}^{\iota},\boldsymbol{R}_0^{\iota},\boldsymbol{t}^{\iota},\boldsymbol{r}^{\iota},\{\boldsymbol{u}_{w,i}^{\iota}\},\boldsymbol{\rho}^{\iota},\boldsymbol{\upsilon}^{\iota}$, and set $\iota=0$.
					\REPEAT		
					\STATE Update ${\rho}_k^{\iota+1} $ via~\eqref{rho_update};
					\STATE Update 	$\boldsymbol{\upsilon}^{\iota+1}$ via~\eqref{v};
					\STATE Obtain 	$\tilde{\boldsymbol{W}}^{\iota+1}$ and $\tilde{\boldsymbol{R}}_0^{\iota+1}$ by solving  Problem in~\eqref{Problem2}, and construct $\bar{\boldsymbol{W}}^{\iota+1}$ and $\bar{\boldsymbol{R}}_0^{\iota+1}$ via~\eqref{construct};
					\STATE Update $\boldsymbol{t}^{\iota+1}$  via the PGD algorithm;
					\STATE Update $\boldsymbol{r}^{\iota+1}$  by \textit{Algorithm 1};
					\STATE Update $\{\boldsymbol{u}_{w,i}^{\iota+1}\}$  via eigenvalue decomposition;
					\STATE Let $\iota = \iota+1$;
					\UNTIL Exit conditions are met.
					\RETURN  $\boldsymbol{W}^\star,\boldsymbol{R}_0^\star,\boldsymbol{t}^\star,\boldsymbol{r}^\star$, and $\{\boldsymbol{u}_{w,i}^\star\}$.
				\end{algorithmic} 
			\end{algorithm}
		\end{small}
		\section{Modification to Colluding Scenarios}
	In this section, we investigate the  transmission covertness against colluding Willies. Unlike the non-colluding case, the covertness constraints in~\eqref{covert2} arising from joint detection  significantly complicate the optimization. However, we  show that the challenging problem can be efficiently solved by incorporating  Woodbury identity in~\cite{petersen2008matrix} and MMSE techniques into the proposed BCD-based optimization framework. In the following, we first reformulate the covertness constraints into a more tractable form, and then detail the necessary algorithmic modifications.
	\vspace{-4mm}
		\subsection{Covertness Constraint Reformulation}
		
		With $	\mathcal{D}(\mathbb{P}_{F,0}||\mathbb{P}_{F,1}) = M(\text{Tr}(\boldsymbol{\Lambda}_1^{-1}\boldsymbol{\Lambda}_0)-\ln\det(\boldsymbol{\Lambda}_1^{-1}\boldsymbol{\Lambda}_0)-$\\$W)$, the Woodbury's formula is employed on the first term, i.e., $(\boldsymbol{A}+\boldsymbol{BC})^{-1} = \boldsymbol{A}^{-1}- \boldsymbol{A}^{-1}\boldsymbol{B}(\boldsymbol{I}+\boldsymbol{C}\boldsymbol{A}^{-1}\boldsymbol{B})^{-1}\boldsymbol{CA}^{-1}$, and thus we can transform  $	\mathcal{D}(\mathbb{P}_{F,0}||\mathbb{P}_{F,1})$ into a more tractable form in~\eqref{Woodbury}, shown at the top of next page.  Note that the procedure (a) holds true since $\mathcal{Q}(\boldsymbol{W},\boldsymbol{R}_0,\boldsymbol{t})\le0.$
			\begin{figure*}[tp]
				\begin{small}
					\begin{subequations}\label{Woodbury}
						\begin{alignat}{2}
							&\mathcal{D}(\mathbb{P}_{F,0}||\mathbb{P}_{F,1}) =\notag\\
							&  M\Big(\text{Tr}\Big((\boldsymbol{\Lambda}_0^{-1}-\boldsymbol{\Lambda}_0^{-1}\boldsymbol{H}_F(\boldsymbol{t})\boldsymbol{W}(\boldsymbol{I}_W+\boldsymbol{W}^H\boldsymbol{H}_F(\boldsymbol{t})^H\boldsymbol{\Lambda}_0^{-1}\boldsymbol{H}_F(\boldsymbol{t})\boldsymbol{W})^{-1}
							\boldsymbol{W}^H\boldsymbol{H}_F(\boldsymbol{t})^H\boldsymbol{\Lambda}_0^{-1})\boldsymbol{\Lambda}_0\Big)-\ln\det(\boldsymbol{\Lambda}_1^{-1}\boldsymbol{\Lambda}_0)-W\Big)\\
							& = M\Big(\underbrace{-\text{Tr}\Big(\boldsymbol{W}^H\boldsymbol{H}_F(\boldsymbol{t})^H\boldsymbol{\Lambda}_0^{-1}\boldsymbol{H}_F(\boldsymbol{t})\boldsymbol{W}(\boldsymbol{I}_W+\boldsymbol{W}^H\boldsymbol{H}_F(\boldsymbol{t})^H\boldsymbol{\Lambda}_0^{-1}
								\boldsymbol{H}_F(\boldsymbol{t})\boldsymbol{W})^{-1}\Big)}_{\triangleq \mathcal{Q}(\boldsymbol{W},\boldsymbol{R}_0,\boldsymbol{t})}-\ln\det(\boldsymbol{\Lambda}_1^{-1}\boldsymbol{\Lambda}_0)\Big)\\
							&\overset{(a)}{\le} M\Big(\ln\det\boldsymbol{\Lambda}_1-\ln\det\boldsymbol{\Lambda}_0\Big).
						\end{alignat} 
					\end{subequations} 
				\end{small}
			\hrule
			\vspace{-2mm}
		\end{figure*}
		Consequently, with $\boldsymbol{\Lambda}_1 = \boldsymbol{\Lambda}_0 + \boldsymbol{H}_F(\boldsymbol{t})\boldsymbol{WW}^H\boldsymbol{H}_F(\boldsymbol{t})^H$ and $\log_2(x) = \frac{\ln (x)}{\ln 2}$, the constraint in~\eqref{covert2} can be approximated as
			\begin{align}\label{covert_colluding_app}
				&\underbrace{\log_2\det\Big(\boldsymbol{I}_W+\frac{1}{\sigma_0^2}\boldsymbol{H}_F(\boldsymbol{t})\boldsymbol{R}_0\boldsymbol{H}_F(\boldsymbol{t})^H\Big)}_{\triangleq~\mathcal{C}_1(\boldsymbol{R}_0,\boldsymbol{t})}\notag\\
				&\underbrace{-\log_2\det\Big(\boldsymbol{I}_W+\frac{1}{\sigma_0^2}\boldsymbol{H}_F(\boldsymbol{t})\boldsymbol{R}_X^1\boldsymbol{H}_F(\boldsymbol{t})^H\Big)}_{\triangleq~\mathcal{C}_2(\boldsymbol{W},\boldsymbol{R}_0,\boldsymbol{t})}\ge\frac{-2\epsilon_F^2}{M\ln2},
			\end{align}
which is still non-convex w.r.t. $\boldsymbol{R}_0$, $\boldsymbol{W}$, and $\boldsymbol{t}$. Fortunately, it is observed that the constraint in~\eqref{covert_colluding_app} is formulated as the difference between channel capacities under different hypotheses. Therefore, we rewrite $\mathcal{C}_1(\boldsymbol{R}_0,\boldsymbol{t})$ and $\mathcal{C}_2(\boldsymbol{W},\boldsymbol{R}_0,\boldsymbol{t})$ into equivalent but more favourable forms, respectively. In terms of $\mathcal{C}_1(\boldsymbol{R}_0,\boldsymbol{t})$, we adopt the MMSE-based  transformation method~\cite{7018097}. Specifically, by introducing an auxiliary matrix $\boldsymbol{U}_1\in\mathbb{C}^{W\times N}$, the mean square error matrix is given by
\begin{align}
	\boldsymbol{E}_1(\boldsymbol{R}_E,&\boldsymbol{U}_1,\boldsymbol{t}) \notag\\ 
	&\triangleq\big(\boldsymbol{U}_1^H\boldsymbol{H}_F(\boldsymbol{t})\boldsymbol{R}_E-\boldsymbol{I}_N\big)\big(\boldsymbol{U}_1^H\boldsymbol{H}_F(\boldsymbol{t})\boldsymbol{R}_E-\boldsymbol{I}_N\big)^H\notag\\
	&+\sigma_0^2\boldsymbol{U}_1^H\boldsymbol{U}_1,
\end{align}
where 
 $\boldsymbol{R}_0=\boldsymbol{R}_E\boldsymbol{R}_E^H$ with $\boldsymbol{R}_E\succeq 0$. In addition, we further introduce an auxiliary positive semidefinite  matrix $\boldsymbol{P}_1\in\mathbb{C}^{N\times N}$, $\boldsymbol{P}_1\succeq 0$. According to~\cite{5756489}, $\mathcal{C}_1(\boldsymbol{R}_0,\boldsymbol{t})$ can be rewritten as 
		\begin{align}
			\mathcal{C}_1(\boldsymbol{R}_0,\boldsymbol{t}) &= \underset{\boldsymbol{U}_1,\boldsymbol{P}_1}{\max}~\mathcal{T}_1(\boldsymbol{R}_E,\boldsymbol{U}_1,\boldsymbol{P}_1,\boldsymbol{t})\notag\\
			&\triangleq \underset{\boldsymbol{U}_1,\boldsymbol{P}_1}{\max} \log_2\det\boldsymbol{P}_1-\text{Tr}\Big(\boldsymbol{P}_1\boldsymbol{E}_1(\boldsymbol{R}_E,\boldsymbol{U}_1,\boldsymbol{t})\Big) + N,
		\end{align}
		and the optimal solutions  of $\boldsymbol{U}_1$ and  $\boldsymbol{P}_1$ can be respectively given by
		\begin{align}
			\boldsymbol{U}_1^\star &= \arg \underset{\boldsymbol{U}_1}{\max}~ \mathcal{T}_1(\boldsymbol{R}_E,\boldsymbol{U}_1,\boldsymbol{P}_1,\boldsymbol{t})\notag\\
			& =  \Big(\sigma_0^2\boldsymbol{I}_W+\boldsymbol{H}_F(\boldsymbol{t})\boldsymbol{R}_E\boldsymbol{R}_E^H\boldsymbol{H}_F(\boldsymbol{t})^H\Big)^{-1}\boldsymbol{H}_F(\boldsymbol{t})\boldsymbol{R}_E,\\
			\boldsymbol{P}_1^\star& = \arg \underset{\boldsymbol{P}_1}{\max}~ \mathcal{T}_1(\boldsymbol{R}_E,\boldsymbol{U}_1,\boldsymbol{P}_1,\boldsymbol{t})=[\boldsymbol{E}_1(\boldsymbol{R}_E,\boldsymbol{U}_1^\star,\boldsymbol{t})]^{-1}.
		\end{align}
	Next, for $\mathcal{C}_2(\boldsymbol{W},\boldsymbol{R}_0,\boldsymbol{t})$, we introduce the following lemma~\cite{5962875}.

	\textit{\textbf{Lemma 1:}} Let $\boldsymbol{T} \in \mathbb{C}^{N\times N}$ be any matrix such that $\boldsymbol{T}\succeq 0$. With the function $f(\boldsymbol{Q})=-\text{Tr}\Big(\boldsymbol{QT}\Big)+\log_2\det\boldsymbol{Q}+N$, we have 
	\begin{equation}
		-\log_2\det\boldsymbol{T} =\underset{\boldsymbol{Q}\succeq 0}{\max}~f(\boldsymbol{Q}),
	\end{equation}
	and the optimal solution is given by $\boldsymbol{Q}^\star = \boldsymbol{T}^{-1}.$
	
	According to Lemma 1, we define that $\boldsymbol{E}_2(\boldsymbol{W},\boldsymbol{R}_E,\boldsymbol{t}) \triangleq\boldsymbol{I}_W+\frac{1}{\sigma_0^2}\boldsymbol{H}_F(\boldsymbol{t})\boldsymbol{R}_X^1\boldsymbol{H}_F(\boldsymbol{t})^H$, and introduce an auxiliary positive semidefinite matrix $\boldsymbol{P}_2\in\mathbb{C}^{W\times W}$, $\boldsymbol{P}_2\succeq 0$. Then,  $\mathcal{C}_2(\boldsymbol{W},\boldsymbol{R}_0,\boldsymbol{t})$ can be formulated as 
	\begin{align}
			\mathcal{C}_2(&\boldsymbol{W},\boldsymbol{R}_0,\boldsymbol{t}) = \underset{\boldsymbol{P}_2}{\max}~\mathcal{T}_2(\boldsymbol{W},\boldsymbol{R}_E,\boldsymbol{P}_2,\boldsymbol{t})\notag\\
		&\triangleq \underset{\boldsymbol{P}_2}{\max} ~\log\det\boldsymbol{P}_2-\text{Tr}\Big(\boldsymbol{P}_2\boldsymbol{E}_2(\boldsymbol{W},\boldsymbol{R}_E,\boldsymbol{t})\Big)+W,
	\end{align}
		where the optimal $\boldsymbol{P}_2$ is given by
		\begin{equation}
			\boldsymbol{P}_2^\star = [\boldsymbol{E}_2(\boldsymbol{W},\boldsymbol{R}_E,\boldsymbol{t})]^{-1}.
		\end{equation}
	Define that $\mathcal{R}(\boldsymbol{R}_E,\boldsymbol{W},\boldsymbol{U}_1,\boldsymbol{P}_1,\boldsymbol{P}_2,\boldsymbol{t})\triangleq \mathcal{T}_1(\boldsymbol{R}_E,\boldsymbol{U}_1,\boldsymbol{P}_1,\boldsymbol{t})+\mathcal{T}_2(\boldsymbol{W},\boldsymbol{R}_E,\boldsymbol{P}_2,\boldsymbol{t})$. With the above derivation, the constraints in~\eqref{covert2} can be approximated as 
	\begin{equation}\label{colluding_constraint}
	\mathcal{R}(\boldsymbol{R}_E,\boldsymbol{W},\boldsymbol{U}_1,\boldsymbol{P}_1,\boldsymbol{P}_2,\boldsymbol{t})\ge\frac{-2\epsilon_F^2}{M\ln2}.
	\end{equation} 
	Thus, the optimization problem with  colluding Willies can be given by
	\begin{subequations}\label{P9}
		\begin{alignat}{2}
			&\underset{ \boldsymbol{W}, \boldsymbol{R}_0,  {\boldsymbol{r}},{\boldsymbol{t},\{\boldsymbol{u}_{w,i}\}}}{\max} \mathcal{F}_2(\boldsymbol{W},\boldsymbol{R}_0,\boldsymbol{t},\boldsymbol{\upsilon})\\
			&\,\text {s.t.}~ \eqref{P1_SNR},\eqref{P1_region},\eqref{P1_distance},\eqref{P1_Power},\eqref{P1_R0},\eqref{colluding_constraint}.
		\end{alignat} 
	\end{subequations} 
It is observed that the optimization problem in~\eqref{P9} differs from that in non-colluding scenarios only in the  constraint in~\eqref{colluding_constraint}. Therefore, we exhibit modifications on the BCD algorithm. The iteration order of  optimization variables in a single loop in \textit{Algorithm 2} is adjusted to: $\boldsymbol{\rho}\rightarrow\boldsymbol{\upsilon}\rightarrow\boldsymbol{U}_1\rightarrow\boldsymbol{P}_1\rightarrow\boldsymbol{P}_2\rightarrow\boldsymbol{W}~\text{and}~\boldsymbol{R}_0\rightarrow\boldsymbol{t}\rightarrow\boldsymbol{r}\rightarrow\{\boldsymbol{u}_{w,i}\}$. Note that the constraint in~\eqref{colluding_constraint} couples beamforming matrices ($\boldsymbol{W}$ and $\boldsymbol{R}_0$) with the transmit APV $\boldsymbol{t}$. Consequently, we focus on the  specific modifications for optimizing these variables, while omitting the remaining steps that are identical to those in Section III for brevity.
	\subsection{Updating Transmit Beamforming}
	With other variables being fixed, the SDP problem in~\eqref{Problem2} is revised  as 
\begin{small}
	\begin{subequations}\label{Problem7}
		\begin{alignat}{2}
			&\underset{ \{\boldsymbol{R}_k\}_{k=1}^{K},\boldsymbol{R}_E}{\max} ~\sum_{k=1}^{K}\Big\{2(1+\rho_k)\upsilon_k\sqrt{\boldsymbol{h}_k^H({\boldsymbol{t}}){\boldsymbol{R}}_k\boldsymbol{h}_k({\boldsymbol{t}})}\notag\\
			&\quad-(1+\rho_k)|\upsilon_k|^2\Big(\boldsymbol{h}_k^H(\boldsymbol{t})(\sum_{c=1}^{K}\boldsymbol{R}_c+\boldsymbol{R}_E\boldsymbol{R}_E^H)\boldsymbol{h}_k(\boldsymbol{t})+\sigma_k^2\Big)\Big\} \label{P7_OBJ}\\
			&\,\text {s.t.}~
			|\alpha_w|^2\sum_{k=1}^{K}\text{Tr}\Big(\boldsymbol{R}_k\boldsymbol{\Xi}_{w,w}^1\Big)+|\alpha_w|^2\text{Tr}\Big(\boldsymbol{R}_E^H\boldsymbol{\Xi}_{w,w}^1\boldsymbol{R}_E\Big)\ge\notag\\
			&\Gamma_w\sum_{c\neq w}^{W}\Big(|\alpha_c|^2\text{Tr}(\boldsymbol{R}_X^1\boldsymbol{\Xi}_{w,c})\Big)+\sigma_r^2\Gamma_w\boldsymbol{u}_{w,1}^H\boldsymbol{u}_{w,1},\forall w\in\mathcal{W},\label{P7_SINR_H1}\\
			&\hphantom{s.t.~}|\alpha_w|^2\text{Tr}\Big(\boldsymbol{R}_E^H\boldsymbol{\Xi}_{w,w}^0\boldsymbol{R}_E\Big)-\sigma_r^2\Gamma_w\boldsymbol{u}_{w,0}^H\boldsymbol{u}_{w,0}\ge\notag\\
			&~~~~~~~~~~~~~~~~~~~~\Gamma_w\sum_{c\neq w}^{W}|\alpha_c|^2\text{Tr}\Big(\boldsymbol{R}_E^H\boldsymbol{\Xi}_{w,c}^0\boldsymbol{R}_E\Big),\forall w \in \mathcal{W},\label{P7_SINR_H0}\\
			&\hphantom {s.t.~}\mathcal{R}(\boldsymbol{R}_E,\boldsymbol{W})\ge\frac{-2\epsilon_F^2}{M\ln2},\label{P7_covert}\\
			&\hphantom {s.t.~}\boldsymbol{R}_E\succeq0,~\sum_{k=1}^{K}\text{Tr}\Big(\boldsymbol{R}_k\Big)+\text{Tr}\Big(\boldsymbol{R}_E^H\boldsymbol{R}_E\Big)\le P_t,\label{P7_power} \\
			&\hphantom {s.t.~} \boldsymbol{R}_k \succeq 0,~\text{rank}(\boldsymbol{R}_k) = 1,1\le k \le K,\label{P7_rank}
		\end{alignat} 
	\end{subequations} 
\end{small}where $\boldsymbol{R}_k = \boldsymbol{w}_k\boldsymbol{w}_k^H$, and $\boldsymbol{R}_X^1 = \sum_{k=1}^{K}\boldsymbol{R}_k + \boldsymbol{R}_E\boldsymbol{R}_E^H$. Here, the constraints in~\eqref{P7_SINR_H1} and~\eqref{P7_SINR_H0} denote the radar SINR constraints under $\mathcal{H}_1$ and $\mathcal{H}_0$, respectively. Unfortunately, after dropping the rank-one constraints in~\eqref{P7_rank}, the optimization problem in~\eqref{Problem7} is still intractable due to the non-concavity of matrix quadratic terms on the left-hand-side (LHS) of~\eqref{P7_SINR_H1} and~\eqref{P7_SINR_H0}, i.e., $\text{Tr}\Big(\boldsymbol{R}_E^H\boldsymbol{\Xi}_{w,w}^i\boldsymbol{R}_E\Big),\forall w \in \mathcal{W}, \forall i \in\{0,1\}$. To deal with this issue, the SCA method can be applied. Specifically, we replace  $\text{Tr}\Big(\boldsymbol{R}_E^H\boldsymbol{\Xi}_{w,w}^i\boldsymbol{R}_E\Big)$ with its first-order Taylor expansion, i.e.,
\begin{align*}
	\text{Tr}&\big(\boldsymbol{R}_E^H\boldsymbol{\Xi}_{w,w}^i\boldsymbol{R}_E\big) \ge \text{Tr}\big(\boldsymbol{R}_E^{l,H}\boldsymbol{\Xi}_{w,w}^i\boldsymbol{R}_E^{l}\big) +\notag\\ &2\Re\big\{\text{Tr}\big(\boldsymbol{R}_E^{l,H}\boldsymbol{\Xi}_{w,w}^i(\boldsymbol{R}_E-\boldsymbol{R}_E^l)\big)\big\},\forall w \in \mathcal{W},\forall i \in \{0,1\}.
\end{align*}
Here, $\boldsymbol{R}_E^l$ is the obtained solution in the $l$-th iteration. By approximating $\text{Tr}\Big(\boldsymbol{R}_E^H\boldsymbol{\Xi}_{w,w}^i\boldsymbol{R}_E\Big)$, the problem in~\eqref{Problem7} is convex and can be solved by using the CVX tool. Note that the construction method in~\eqref{construct} can be directly employed for the SDP problem in~\eqref{Problem7}, guaranteeing optimality under $\mathcal{H}_1.$ The proof follows a similar procedure to that in Appendix B, and is therefore omitted.

\subsection{Updating Transmit Antenna Placement}
For transmitting antenna positions $\boldsymbol{t}$, we only need to modify the Step.~2 of the PGD procedure in~\eqref{b}. The sole distinction lies in different covertness constraints. To address this issue, we employ the second-order Taylor expansion theorem to obtain a tractable lower bound of $\mathcal{R}(\boldsymbol{t})$, i.e., $\mathcal{R}(\boldsymbol{t})\geq\mathcal{R}(\boldsymbol{t^l}) + \nabla\mathcal{R}(\boldsymbol{t^l})^T(\boldsymbol{t} - \boldsymbol{t}^l)-\frac{\omega}{2}||{\boldsymbol{t}}-{\boldsymbol{t}}^{l}||^2_2.$
Note that a positive real number $\omega$ is selected to satisfy $\omega\boldsymbol{I}_N\succeq\nabla^2\mathcal{R}(\boldsymbol{t})$, with $\nabla^2\mathcal{R}(\boldsymbol{t})\in\mathbb{C}^{N\times N}$ being the Hessian matrix. The construction of  $\omega$ is similar to that in Appendix D and is omitted to avoid redundancy. Then, the problem in Step.~2 in~\eqref{b} can be revised as 
\begin{small}
	\begin{subequations}\label{Problem8}
		\begin{alignat}{2}
			&\underset{{\boldsymbol{t}}}{\min} \quad ||\boldsymbol{t}-\boldsymbol{m}^{l+1}||_2^2 \\
			&\,\text {s.t.}~\mathcal{R}(\boldsymbol{t^l}) + \nabla\mathcal{R}(\boldsymbol{t^l})^T(\boldsymbol{t} - \boldsymbol{t}^l)-\frac{\omega}{2}||{\boldsymbol{t}}-{\boldsymbol{t}}^{l}||^2_2\ge \frac{-2\epsilon_F^2}{M\ln2},\\
			&\hphantom {s.t.~}   \eqref{P5_distance},\eqref{P5_SINR},\eqref{P5_covert},
		\end{alignat} 
	\end{subequations} 
\end{small}
which is convex and can be solved by existing solvers~\cite{grant2008cvx}. 

%Overall, the proposed BCD method in Section III can be modified to guarantee transmission covertness against colluding Willies. In particular, the covertness constraints in~\eqref{covert_noncolluding} are replaced by~\eqref{covert_colluding_app}. Then, an MMSE-based method along with SCA is proposed to facilitate system optimization. 
\vspace{-4mm}
\subsection{Convergence and Complexity Analysis}
For both non-colluding and colluding Willies, each variable is updated with either an optimal or a locally optimal solution.  According to~\cite{boyd2004convex}, the unified algorithm  asymptotically converges to a  stationary  point of~\eqref{Problem1}. The computational complexity  is presented in~Table~\ref{table1}, shown at the top of this page. Note that the  complexity under colluding Willies does not substantially increase  compared to that under non-colluding scenarios.  This is because, although the colluding-related constraints introduce additional quadratic matrix forms, the overall number of covertness constraints is reduced. This trade-off ensures that the computational overhead remains relatively invariant.
	\begin{table*}[t]
		\centering
		\caption{Complexity of the Proposed Algorithm.}\label{table1}
		\renewcommand{\arraystretch}{2.0}
		\begin{tabular}{c|c|c}
			\hline
			Optimization Variable &  Non-colluding Willies & Colluding Willies\\
			\hline
			$\boldsymbol{\rho}$ &\multicolumn{2}{c}{$\mathcal{O}(K^2N+KN^2)$}  \\
			\hline
			$\boldsymbol{\upsilon}$ &\multicolumn{2}{c}{$\mathcal{O}(KN^2)$}  \\
			\hline
			$\boldsymbol{U}_1$ & -- & $\mathcal{O}(N^3 + W^3 + N^2W + W^2N)$ \\ 
			\hline
			$\boldsymbol{P}_1$ & -- & $\mathcal{O}(N^3 + N^2 W)$\\
			\hline
			$\boldsymbol{P}_2$ & -- & $\mathcal{O}(W^3 + WN^2 + NW^2)$\\
			\hline
			$\boldsymbol{W}$ and $\boldsymbol{R}_0 (\boldsymbol{R}_E)$ & $\mathcal{O}(\sqrt{2W+NK}(N^6K^3+3WN^4K^2))$ & $\mathcal{O}(\sqrt{4W+KN}(N^6K^3+2WN^6K))$\\
			\hline
			$\boldsymbol{t}$ & $\mathcal{O}(\sqrt{6W+N}WN^3)$ & $\mathcal{O}(\sqrt{4W+N}WN^3)$\\
			\hline
			$\boldsymbol{r}$ & \multicolumn{2}{c}{$\mathcal{O}(\sqrt{4W+N}WN^3)$}\\
			\hline
			$\{\boldsymbol{u}_{w,i}\}$ & \multicolumn{2}{c}{$\mathcal{O}(N^3)$}\\
			\hline
		\end{tabular}
	\end{table*}
	
		\section{Simulation Results}
		In this section, computer simulations are conducted to  evaluate the performance of the proposed method. We compare our scheme with three baseline schemes: \textbf{1) Upper bound performance scheme}: The optimization is performed to maximize the sum rate without covertness constraints; \textbf{2) Fixed position antenna (FPA)}: The BS is equipped with uniform linear arrays, with $N$ transmitting/receiving antennas spaced between intervals of $\frac{\lambda}{2}$;   \textbf{3) Greedy antenna selection (GAS)}: The moving regions are quantized into discrete ports spaced by $\frac{\lambda}{2}$. The greedy algorithm is employed for the optimization on antenna positions\cite{yang2025robust}.
%		\captionsetup{font={small},labelsep=period}
%		\captionsetup[subfloat]{font=small}
%		
%		\begin{figure}
%			\centering 
%			\subfloat[]{ %
%				\begin{minipage}[t]{0.5\linewidth}%
%					\centering \includegraphics[width=1.5in]{iteration.eps}\label{iteraion} 
%					%
%			\end{minipage}}
%			\subfloat[]{ %
%				\begin{minipage}[t]{0.5\linewidth}%
%					\centering \includegraphics[width=1.5in]{Pt_revised.eps} \label{Pt}
%					%
%			\end{minipage}}
%			\caption{(a) Convergence behavior.  (b) Covert sum rate versus transmit power $P_t$.}\label{SSS1}
%			\vspace{-4mm}
%		\end{figure}
%		\begin{figure}
%			\centering 
%			\subfloat[]{ %
%				\begin{minipage}[t]{0.5\linewidth}%
%					\centering \includegraphics[width=1.5in]{SNR_revised.eps}\label{SNR}
%					%
%			\end{minipage}}
%			\subfloat[]{ %
%				\begin{minipage}[t]{0.5\linewidth}%
%					\centering 
%					\includegraphics[width=1.5in]{epsilon.eps} \label{epsilon}
%					%
%			\end{minipage}}
%			\caption{(a) Trade-off between covert sum rate and radar SNR $\Gamma$. (b) Covert sum rate versus covertness level $\epsilon$.}
%			\label{SSS2}
%			\vspace{-6mm}
%		\end{figure}
		
		In our simulation, we assume that the BS is located at~(0,~0) m. The users are randomly distributed in a circle centered at (40,~0) m with a radius of 5 m. The numbers of transmitting and receiving paths are identical, i.e., $L_k = L = 12,\forall k$.  The PRM  is given by  $\boldsymbol{\Sigma}_k = \text{diag}\{\sigma_{k,1},\dots,\sigma_{k,L}\}$ with $\sigma_{k,l}\sim\mathcal{CN}\left(0,\frac{c_k^2}{L}\right)$. Note that $c_k^2 = C_0d_k^{-a}$ denotes the large-scale path loss, where $C_0 = -30~\text{dB}$, and the path-loss exponents $\alpha$ for communcation and sensing links are set to  3.2 and 2.6, respectively. Unless otherwise specified, the number of sensing targets are set as $W=2$, and the elevation angles are set to be $20\degree$ and $105\degree$, respectively.  Other parameters: $K=3,N=6,M=32, \Gamma_w = \Gamma = 10~\text{dB},\epsilon_w = \epsilon_F=\epsilon=0.05,\forall w,\lambda = 0.1~\text{m},d=\frac{\lambda}{2}$, and $D = 14\lambda.$
			\begin{figure}[t]
			\centering
			\includegraphics[width=0.40\textwidth]{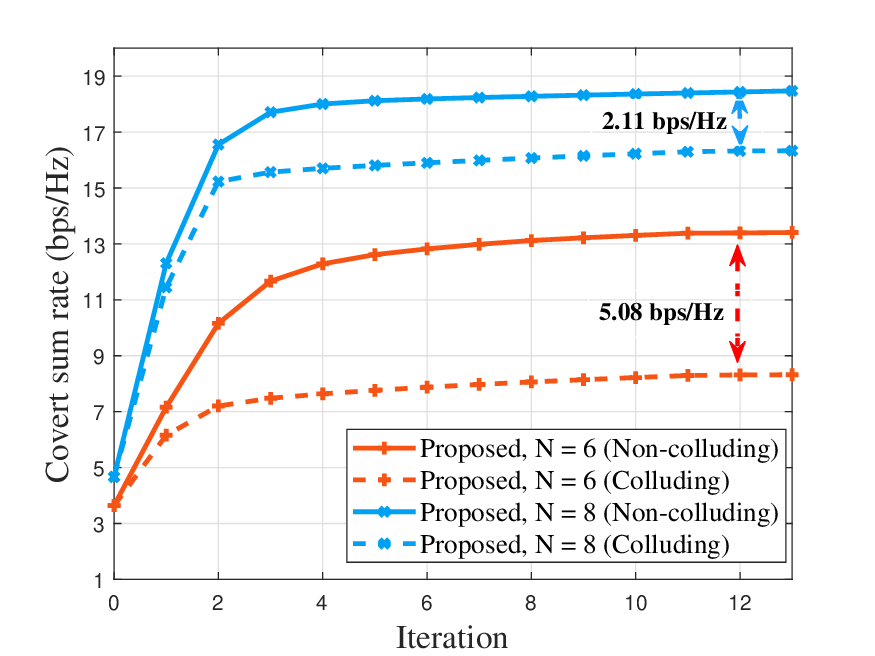}
			\captionsetup{font={normalsize},labelsep=period,singlelinecheck=off}
			\caption{Convergence behaviour of the BCD algorithm.} 
			\label{Convergence} 
		\end{figure}%
		\begin{figure}[t]
			\centering
			\includegraphics[width=0.40\textwidth]{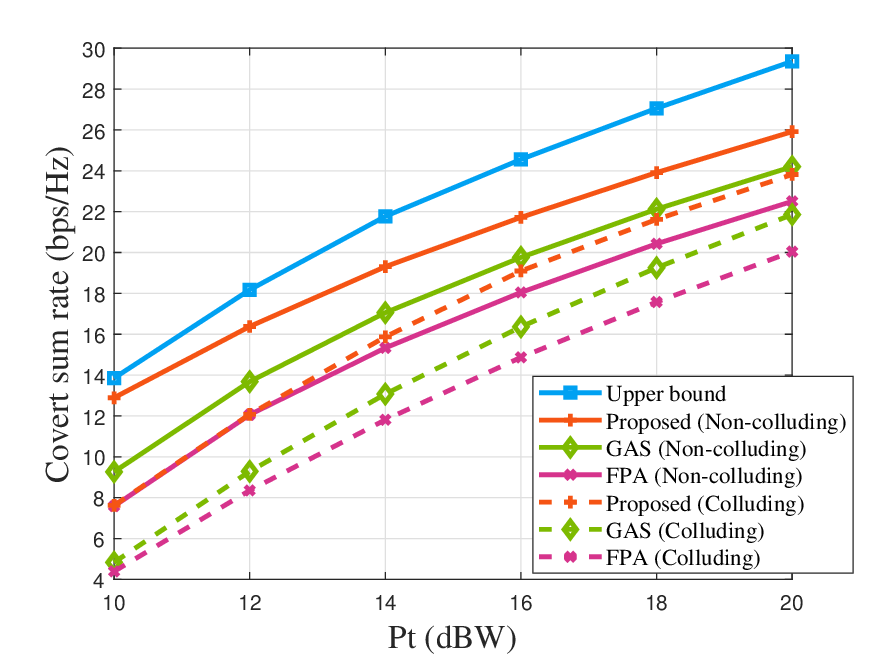}
			\captionsetup{font={normalsize},labelsep=period,singlelinecheck=off}
			\caption{The covert sum rate versus transmission power $P_t$.} 
			\label{Pt} 
		\end{figure}%
			\begin{figure}[t]
			\centering
			\includegraphics[width=0.40\textwidth]{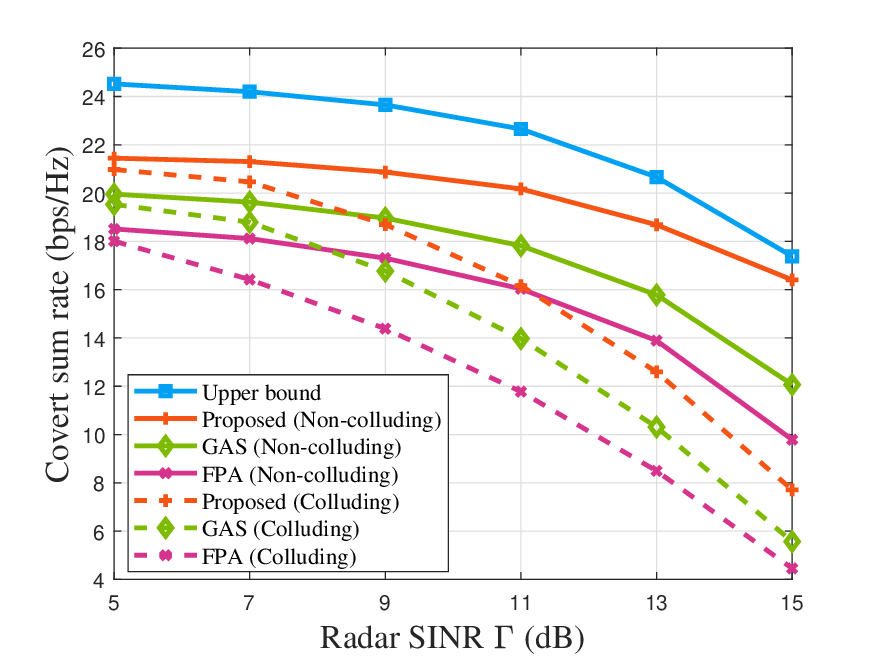}
			\captionsetup{font={normalsize},labelsep=period,singlelinecheck=off}
			\caption{The covert sum rate versus radar SINR $\Gamma$.} 
			\label{SINR} 
		\end{figure}%
			\begin{figure}[t]
			\centering
			\includegraphics[width=0.40\textwidth]{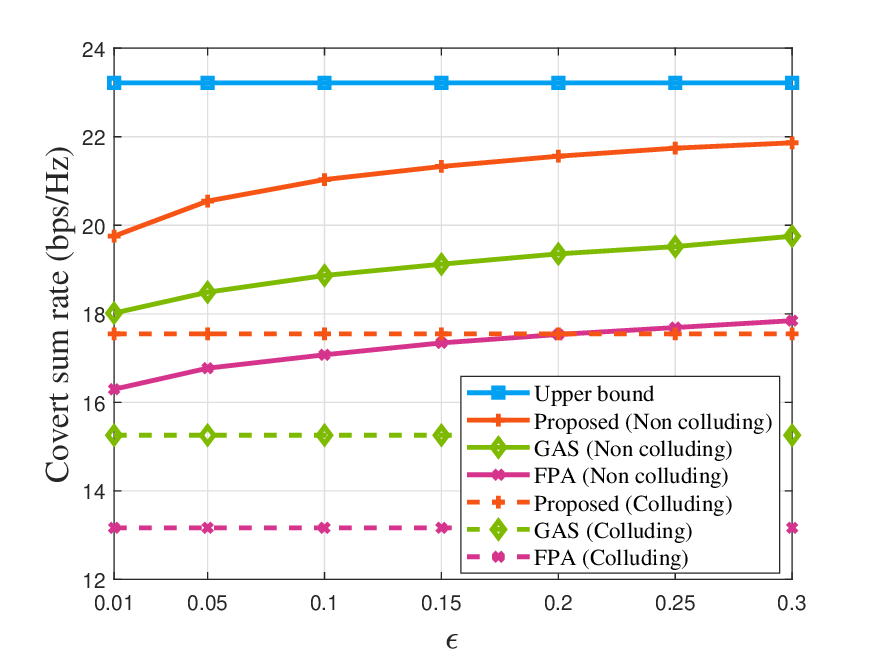}
			\captionsetup{font={normalsize},labelsep=period,singlelinecheck=off}
			\caption{The covert sum rate versus covertness level $\epsilon$.} 
			\label{epsilon} 
		\end{figure}%
			\begin{figure}[t]
			\centering
			\includegraphics[width=0.40\textwidth]{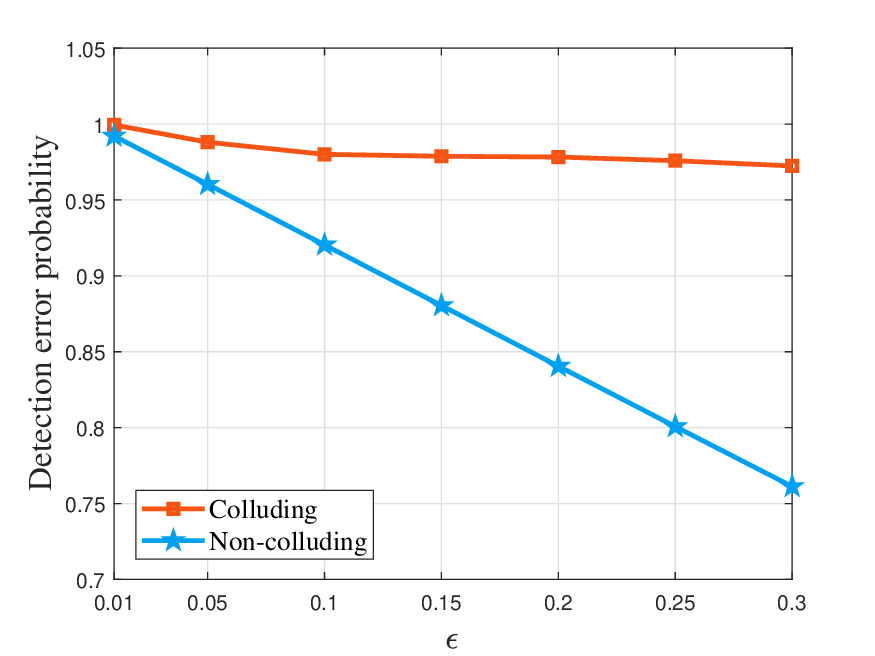}
			\captionsetup{font={normalsize},labelsep=period,singlelinecheck=off}
			\caption{The DEP versus covertness level $\epsilon$.} 
			\label{CDF} 
		\end{figure}%
			\begin{figure}[t]
			\centering
			\includegraphics[width=0.40\textwidth]{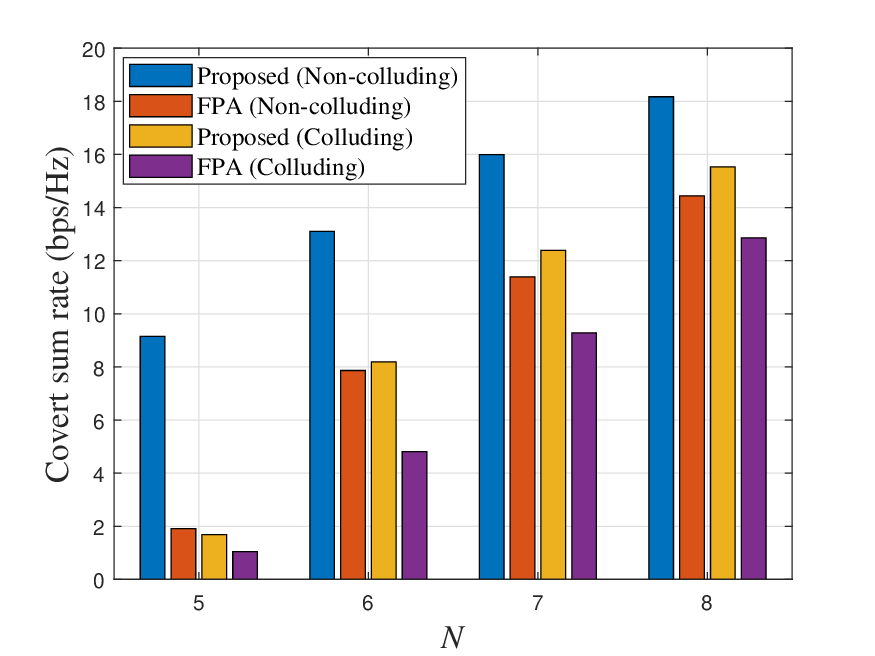}
			\captionsetup{font={normalsize},labelsep=period,singlelinecheck=off}
			\caption{The covert sum rate versus antenna number  $N$.} 
			\label{N} 
		\end{figure}%
			\begin{figure}[t]
			\centering
			\includegraphics[width=0.40\textwidth]{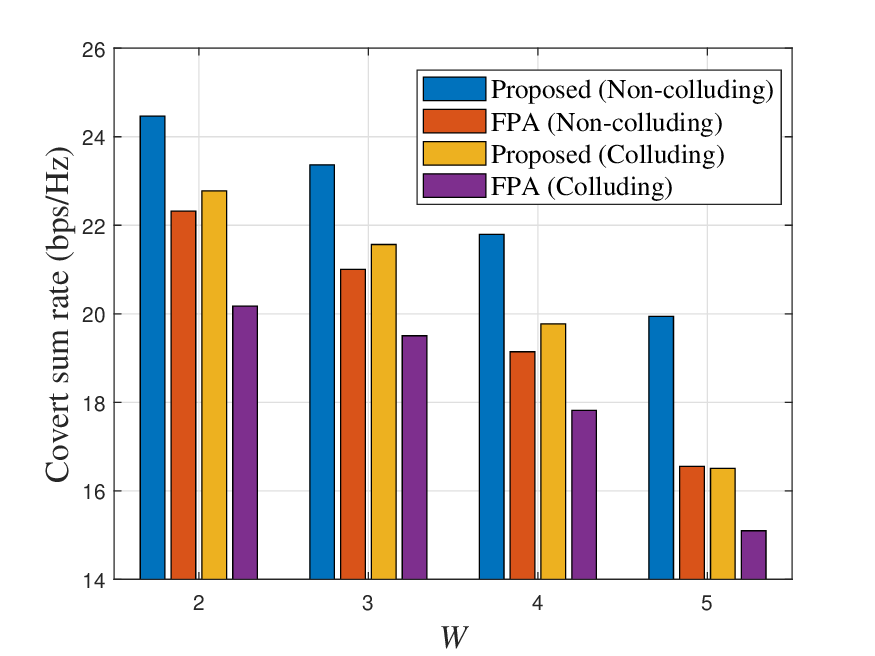}
			\captionsetup{font={normalsize},labelsep=period,singlelinecheck=off}
			\caption{The covert sum rate versus Willie number  $W$.} 
			\label{Willie} 
		\end{figure}%
		%		\begin{figure}[t]
			%			\centering
			%			\includegraphics[width=0.30\textwidth]{iteration.eps}
			%			\captionsetup{font={small},labelsep=period,singlelinecheck=off}
			%			\caption{Convergence behaviour.} 
			%			\label{iteraion} 
			%			\vspace{-5mm}
			%		\end{figure}%
		
		We first present the convergence behavior of \textit{Algorithm 2} in Fig.\,\ref{Convergence}. It can be observed that the covert sum rate under both colluding and non-colluding Willie scenarios monotonically increases and eventually stabilizes as the iterations progress. For instance, with $N = 6$ under colluding Willies, the achievable rate improves from 3.6 bps/Hz to 13.1 bps/Hz, highlighting the effectiveness of the proposed design. In most cases, fewer than 10 iterations are sufficient. Note that the number of antennas $N$ has a minor impact on the convergence rate. This is because many subproblems can be optimally or near-optimally solved (e.g., $\boldsymbol{\rho},\boldsymbol{\upsilon},\boldsymbol{W},\boldsymbol{R}_0,\varrho,\boldsymbol{P}_1,\boldsymbol{P}_2,\boldsymbol{U}_1$, and $\{\boldsymbol{u}_{w,i}\}$), which enables the proposed algorithm to quickly converge to a local optimum. We further define colluding gain as the performance gap between non-colluding and colluding Willie scenarios.	It can be observed that the colluding gain diminishes as the number of MAs increases, falling from 5.08~bps/Hz to 2.11~bps/Hz. This is because the additional spatial DoFs provided by more MAs can be effectively leveraged to counteract the enhanced joint detection capability of colluding Willies.
		
		Fig.\,\ref{Pt} illustrates the covert sum rate versus transmission power $P_t$. It can be observed that the proposed scheme significantly outperforms both the GAS and FPA benchmarks across the entire range of $P_t$. This performance gain is attributed to the active movement of antenna elements, which not only enhances the desired channel conditions for both radar and communication,  but also provides additional spatial DoFs to effectively mitigate interference. Furthermore, the results reveal a notable colluding gain. Specifically, the sum rate achieved in the non-colluding Willies scenario is strictly higher than that of the colluding counterpart.
		This is because for colluding Willies, the joint signal processing significantly enhances the detection capability. Consequently, for a given covertness level~$\epsilon$, the resulting covertness constraint becomes more stringent, thereby restricting the feasible region of optimization. 
		Another interesting observation is that as  $P_t$ increases, the performance gap between the proposed scheme in the non-colluding Willie scenario and the upper bound scheme gradually widens (from 1.0 bps/Hz to 3.4 bps/Hz). This is because constraints in~\eqref{covert_noncolluding} limit the discrepancy in Willies' received signal strength under different hypotheses, which in turn restricts transmit power allocation, leading to the widening performance gap.

		%		\begin{figure}[t]
			%			\centering
			%			\includegraphics[width=0.30\textwidth]{SNR.eps}
			%			\captionsetup{font={small},labelsep=period,singlelinecheck=off}
			%			\caption{Trade-off between covert sum rate and radar SNR $\Gamma$.} 
			%			\label{SNR} 
			%			\vspace{-5mm}
			%		\end{figure}%.
		
		Fig.\,\ref{SINR} demonstrates communication-sensing trade-off by characterizing the covert rate versus radar SINR threshold~$\Gamma$ with $P_t = 15$ dBW. As $\Gamma$ increases, the achievable covert rate of all schemes shows a declining trend. This is reasonable since a larger $\Gamma$ requires more power to radar waveforms. In particular, the proposed scheme outperforms the conventional FPA scheme in both non-colluding and colluding scenarios, which  demonstrates the potential of MA. Notably, the proposed scheme exhibits greater robustness in non-colluding Willie scenarios compared to colluding ones as $\Gamma$ increases. This stems from the dual role of the sensing targets as Willies, where superior radar sensing performance directly imposes a tighter covertness constraint against colluding Willies. Moreover, we note that the performance of the GAS scheme lies between the proposed and FPA benchmark schemes. This is because, in the GAS scheme, only a limited number of antenna ports are available within the moving region. The greedy approach can only find a good feasible solution rather than the optimal one, leading to the performance gap compared to the continuous optimization in the proposed scheme.

		In Fig.\,\ref{epsilon}, we plot the covert sum rate against covertness level $\epsilon$. For non-colluding Willies, when $\epsilon$ becomes larger, the covertness constraint is looser. Consequently, higher  throughput  can be achieved. With $\epsilon = 0.3$, the proposed scheme experiences only about 1.4 bps/Hz compared to the upper bound, demonstrating superiority of the proposed algorithm. However, for colluding Willies,  the covert sum rate exhibits a distinctive insensitivity to the relaxation of the covertness level $\epsilon$.  This phenomenon is due to the fact that the joint signal processing employed by colluding Willies significantly amplifies their detection sensitivity, which greatly compresses the feasible optimization region.  Within this restricted range, the marginal optimization DoFs facilitated by a larger $\epsilon$ is insufficient to overcome the severe interference and noise conditions, resulting in negligible gain in throughput. 
		
			To further corroborate our interpretation,
			we show the DEP versus $\epsilon$ in  Fig.\,\ref{CDF}. The results verify the theoretical analysis that when $\epsilon$ becomes larger, the DEP becomes lower. It is observed that as $\epsilon$ increases, the DEP for colluding Willies does not decrease significantly, which is consistent with the observations  in  Fig.\,\ref{epsilon}. This phenomenon occurs because any slight adjustment in beamforming and antenna placement to boost the covert rate leads to a sharp surge in signal leakage even with a relaxed $\epsilon$.  This inevitably restricts the system's ability to trade covertness for rate improvement, thereby keeping the DEP nearly saturated at its maximum level.
		
		We then plot the covert rate versus antenna number $N$ with $P_t = 10$ dBW in Fig.\,\ref{N}. The results exhibit that the covert sum rate of all schemes increases significantly as $N$ scales up, as a larger $N$ provides more spatial multiplexing gains to balance sensing, communication, and covertness requirements. It is observed that the colluding gain becomes less pronounced as $N$ increases, which is consistent with the observations in Fig. \ref{Convergence}. Notably, the proposed scheme consistently maintains a substantial advantage over the FPA benchmark, especially in the small-$N$ regime, demonstrating that antenna mobility can effectively compensate for limited spatial resources to ensure high-level covertness.
		
		Fig.\,\ref{Willie} simulates the effect of Willie number $W$ on the covert sum rate with $N = 8$ and $P_t = 15$ dBW.  When $W \le 4$, the colluding gain remains below $2$~bps/Hz. However, this gap nearly doubles as $W$ increases to $5$, which validates the potent joint detection capability of colluding Willies.  Nonetheless, the proposed MA-enhanced scheme maintains a more stable performance gain compared with the FPA scheme, demonstrating the effectiveness of flexible antenna movement in enhancing transmission covertness.

		\section{Conclusion}		
		In this paper, we have investigated a movable antenna-enhanced covert DFRC system. A covert sum  rate maximization  problem was formulated by jointly designing transmit beamforming vectors, receiving filter, and transceiver antenna placement. Both colluding and non-colluding Willies were considered. Specifically, for non-colluding Willies, we developed an efficient BCD-based algorithm, incorporating SDR, PGD, SCA, and Dinkelbach transformation methods. For colluding Willies, we first derived the minimum DEP under the optimal  likelihood ratio test. We subsequently proposed an MMSE-incorporated optimization framework. We further provided a comprehensive complexity analysis for the proposed algorithm. Simulation results showed that our method can significantly improve the covert sum rate, and achieve a  satisfactory trade-off between communication and radar performance compared with existing benchmarks. Overall, our work underscores the great potential of MAs, and provides a promising solution for covert-aware 6G wireless networks.

		\appendices
		\section{Proof of Theorem 2}
		Recalling that $\boldsymbol{H}_F(\boldsymbol{t}) = [\beta_1^*\mathbf{a}_t(\varphi_1,{\boldsymbol{t}}),\dots,\beta_W^*\mathbf{a}_t(\varphi_W,{\boldsymbol{t}})]^H$, the received signals $\boldsymbol{Y}_F\in\mathbb{C}^{W\times M}$ in one round can be recast as 
		\begin{equation}
			\boldsymbol{Y}_F = \boldsymbol{H}_F(\boldsymbol{t})\boldsymbol{X}+\boldsymbol{N}_F,
		\end{equation}
		where $\boldsymbol{X} = [\boldsymbol{x}(1),\dots,\boldsymbol{x}(M)]\in\mathbb{C}^{N\times M}$, and $\boldsymbol{N}_F = [\boldsymbol{n}_F(1),\dots,\boldsymbol{n}_F(M)]\in\mathbb{C}^{W\times M}$. Since the transmitted signals $\boldsymbol{X}$ are assumed to be Gaussian distributed, the PDFs of $\boldsymbol{Y}_F$ under different hypotheses can be respectively given by 
		\begin{small}
			\begin{align}
				\mathbb{P}_{F
					,i} &= \mathbb{P}(\boldsymbol{Y}_F|\mathcal{H}_i) \notag\\&= \frac{1}{\pi^{MW}(\det\boldsymbol{\Lambda}_i)^M} \exp\left(-\text{tr}(\boldsymbol{Y}_F^H\boldsymbol{\Lambda}_i^{-1}\boldsymbol{Y}_F)\right),\forall i \in\{0,1\},\label{PDF_YF}
			\end{align}
		\end{small}Based on the Neyman-Pearson criterion, the optimal test to detect the covert transmission is the likelihood ratio test. The log likelihood ratio (LLR) can be given by 
		\begin{align}
			\text{LLR} = &\ln \mathbb{P}(\boldsymbol{Y}_F|\mathcal{H}_1)  - \ln \mathbb{P}(\boldsymbol{Y}_F|\mathcal{H}_0) \notag\\
			 =&\text{Tr}(\boldsymbol{Y}^H_F\left(\boldsymbol{\Lambda}_0^{-1}-\boldsymbol{\Lambda}_1^{-1}\right)\boldsymbol{Y}_F) + M\ln \frac{\det \boldsymbol{\Lambda}_0}{\det \boldsymbol{\Lambda}_1}.
		\end{align}
		Considering the equal prior probabilities of two hypothesis, the optimal test for colluding  Willies can be expressed by
		\begin{equation}
			\text{LLR}\overset{\mathcal{D}_1}{\underset{\mathcal{D}_0}{\gtrless}}0 \Rightarrow ||\boldsymbol{V}^{1/2}\boldsymbol{Y}_F||_F^2\overset{\mathcal{D}_1}{\underset{\mathcal{D}_0}{\gtrless}} M\ln \frac{\det \boldsymbol{\Lambda}_1}{\det \boldsymbol{\Lambda}_0}.
		\end{equation}
		Thus, the result in~\eqref{decision_colluding} can be obtained. 
		
		Next, we derive the minimum DEP for colluding Willies. For brevity, we only show the derivation under hypothesis $\mathcal{H}_0$, and similar derivations can be performed for hypothesis $\mathcal{H}_1$. Specifically, the term $||\boldsymbol{V}^{1/2}\boldsymbol{Y}_F||_F^2$ can be recast as 
		\begin{equation}
			||\boldsymbol{V}^{1/2}\boldsymbol{Y}_F||_F^2 =\sum_{m=1}^{M}\boldsymbol{y}_F(m)^H\boldsymbol{V}\boldsymbol{y}_F(m).
		\end{equation}
		 Recalling that  $\boldsymbol{y}_F(m)\sim \mathcal{CN}(\boldsymbol{0}_{W\times1},\boldsymbol{\Lambda}_0), \forall m,$ we first whiten $\boldsymbol{y}_F(m)$ as  
		 \begin{equation}
		 	\tilde{\boldsymbol{y}}_F(m) = \boldsymbol{\Lambda}_0^{-1/2}\boldsymbol{y}_F(m)\sim\mathcal{CN}(\boldsymbol{0}_{W\times1},\boldsymbol{I}_W).
		 \end{equation}
		 Thus, we obtain  $||\boldsymbol{V}^{1/2}\boldsymbol{Y}_F||_F^2 =\sum_{m=1}^{M}\tilde{\boldsymbol{y}}_F(m)^H\tilde{\boldsymbol{V}}\tilde{\boldsymbol{y}}_F(m)$, where $\tilde{\boldsymbol{V}} =\boldsymbol{\Lambda}_0^{1/2}\boldsymbol{V}\boldsymbol{\Lambda}_0^{1/2}\in\mathbb{C}^{W\times W}$.  Through eigenvalue decomposition, i.e.,   $\tilde{\boldsymbol{V}} = \boldsymbol{U}\boldsymbol{\Theta}_0\boldsymbol{U}^H$, where $\boldsymbol{\Theta}_0 = \text{diag}(\lambda_1^{0},\dots,\lambda_W^{0})\in \mathbb{C}^{W\times W}$ under $\mathcal{H}_0$, we can obtain 
		 \begin{equation*}
		 	||\boldsymbol{V}^{1/2}\boldsymbol{Y}_F||_F^2 = \sum_{m=1}^{M}\boldsymbol{z}(m)^H\boldsymbol{\Theta}_0\boldsymbol{z}(m) = \sum_{m=1}^{M}\sum_{w=1}^{W}\lambda_w^0|z_{m,w}|^2,
		 \end{equation*} 
		 where $\boldsymbol{z}(m) = \boldsymbol{U}^H\tilde{\boldsymbol{y}}_F(m)\sim\mathcal{CN}(\boldsymbol{0}_{W\times1},\boldsymbol{I}_W),$ and $z_{m,w}$ denotes the $w$-th element of $\boldsymbol{z}(m)$. Since $|z_{m,w}|^2$ follows an exponential distribution with rate parameter $\iota_{m,w}= 1$, i.e., $|z_{m,w}|^2\sim \textbf{\text{exp}}(1)$, the term $||\boldsymbol{V}^{1/2}\boldsymbol{Y}_F||_F^2$ follows generalized Erlang   distribution~\cite{693785}. This can be considered as $W$ distinct sums of Erlang-distributed variables with shape parameter and scale parameter being $M$ and $\lambda_w^0$, respectively. Thus, the FAP and MDP can be respectively given by 
		 \begin{small}
		 	 \begin{align}
		 		&\mathbb{P}(\mathcal{D}_1|\mathcal{H}_0) = \text{Pr}\left(||\boldsymbol{V}^{1/2}\boldsymbol{Y}_F||_F^2\ge \chi|\mathcal{H}_0\right) = 1 - F(\chi|\frac{1}{\lambda_1^0},\dots,\frac{1}{\lambda_W^0}),\notag\\
		 		&\mathbb{P}(\mathcal{D}_0|\mathcal{H}_1) = \text{Pr}\left(||\boldsymbol{V}^{1/2}\boldsymbol{Y}_F||_F^2\le \chi|\mathcal{H}_1\right) = F(\chi|\frac{1}{\lambda_1^1},\dots,\frac{1}{\lambda_W^1}).\notag
		 	\end{align}
		 \end{small}
		 
		 Thus, the result in~\eqref{colludingMDP}  is obtained.\hfill$\Box$

		\section{Performance Analysis  of~\eqref{construct}}
		First, one can derive that
		\begin{equation}
			\boldsymbol{h}_k^H\bar{\boldsymbol{R}}_k\boldsymbol{h}_k = \boldsymbol{h}_k^H{\boldsymbol{w}}_k{\boldsymbol{w}}_k^H\boldsymbol{h}_k = \boldsymbol{h}_k^H\tilde{\boldsymbol{R}}_k\boldsymbol{h}_k,1\le k \le K,
		\end{equation} 
		where $\boldsymbol{h}_k$ is used as a shorthand for $\boldsymbol{h}_k(\boldsymbol{t})$.
	 Thus, the value of the objective function  $\mathcal{F}_2(\boldsymbol{W},\boldsymbol{R}_0)$ in~\eqref{F2} remains unchanged. Next, we show that $\tilde{\boldsymbol{R}}_k-\bar{\boldsymbol{R}}_k\succeq 0,1\le k \le K.$ For any $\boldsymbol{v}\in\mathbb{C}^{N\times1}$, it holds that
	 \begin{equation}
	 	\boldsymbol{v}^H(\tilde{\boldsymbol{R}}_k-\bar{\boldsymbol{R}}_k)\boldsymbol{v} = \boldsymbol{v}^H\tilde{\boldsymbol{R}}_k\boldsymbol{v} - (\boldsymbol{h}_k^H\tilde{\boldsymbol{R}}_k\boldsymbol{h}_k)^{-1}|\boldsymbol{v}^H\tilde{\boldsymbol{R}}_k\boldsymbol{h}_k|^2.
	 \end{equation}
		According to the Cauchy-Schwarz inequality, we have 
		\begin{equation}
			(\boldsymbol{h}_k^H\tilde{\boldsymbol{R}}_k\boldsymbol{h}_k)(\boldsymbol{v}^H\tilde{\boldsymbol{R}}_k\boldsymbol{v}) \ge |\boldsymbol{v}^H\tilde{\boldsymbol{R}}_k\boldsymbol{h}_k|^2.
		\end{equation}
		So $\boldsymbol{v}^H(\tilde{\boldsymbol{R}}_k-\bar{\boldsymbol{R}}_k)\boldsymbol{v}\ge 0$ holds true for any $\boldsymbol{v} \in \mathbb{C}^{N\times1}$, i.e., $\tilde{\boldsymbol{R}}_k-\bar{\boldsymbol{R}}_k\succeq 0$. We can conclude that  $\bar{\boldsymbol{R}}_0 = \boldsymbol{R}_X^1-\sum_{k=1}^{K}\bar{\boldsymbol{R}}_k\succeq \tilde{\boldsymbol{R}}_0$. Consequently, with $i=1$, the constraints in~\eqref{P2_radar_SINR},~\eqref{P2_covert},~\eqref{P2_power} and~\eqref{P2_rank} are met. Thus, we can verify that $\{\bar{\boldsymbol{R}}_k\}_{k=0}^{K}$ is a feasible solution, and  furthermore, it is also a global optimum for covert transmission under $\mathcal{H}_1$\footnote{With $\bar{\boldsymbol{R}}_0\succeq\tilde {\boldsymbol{R}}_0$, the constraints in~(\ref{P2_radar_SINR}) under $\mathcal{H}_0$ may not be satisfied, which indicates that the construction in~(\ref{construct}) may deteriorate sensing performance under $\mathcal{H}_0.$ However, our simulation results demonstrate that such a construction method has a negligible impact on the radar SINR. Furthermore, the subsequent optimization on $\boldsymbol{t,r,}$ and $\{\boldsymbol{u}_{w,i}\}$ can be performed to ensure sensing SINR for $\mathcal{H}_0.$ Therefore, we argue that such a heuristic approach remains reasonable.}.
		\section{Derivation of  the Gradient in~\eqref{gradient}}
		Recall that $\boldsymbol{h}_k^H(\boldsymbol{t}) = \boldsymbol{1}_{L_k}^H\boldsymbol{\Sigma}_k\boldsymbol{G}_k(\boldsymbol{t})$,  $\tilde{\mathcal{F}}_{i,j}(\boldsymbol{t})$ can be recast as   $\tilde{\mathcal{F}}_{i,j}(\boldsymbol{t}) = \boldsymbol{a}_i^H\boldsymbol{G}_i(\boldsymbol{t})\boldsymbol{R}_j\boldsymbol{G}_i^H(\boldsymbol{t})\boldsymbol{a}_i$, where $\boldsymbol{a}_i = \boldsymbol{\Sigma}_i^H\boldsymbol{1}_{L_i}\in\mathbb{C}^{L_i\times 1}$. Let us denote the $(n,m)$-th element of $\boldsymbol{R}_j$ as $\boldsymbol{R}_j(n,m) = |\boldsymbol{R}_j(n,m)|e^{\jmath\angle\boldsymbol{R}_j(n,m)}$, and the $l$-th element of $\boldsymbol{a}_i$ as $\boldsymbol{a}_{i}(l) = |\boldsymbol{a}_{i}(l)|e^{\jmath\angle \boldsymbol{a}_{i}(l)}$.
		Thus,  $\tilde{\mathcal{F}}_{i,j}(\boldsymbol{t})$ is recast as 
		\begin{small}
			\begin{align}
				\tilde{\mathcal{F}}_{i,j}(\boldsymbol{t})=& \sum_{n=1}^{N}\sum_{l=1}^{L_i}|\boldsymbol{a}_{i}(l)|^2\boldsymbol{R}_j(n,n)
				+\notag\\
				&\sum_{n=1}^{N}\sum_{l=1}^{L_i-1}\sum_{p=l+1}^{L_i}2\mu_{i,j,n,n,l,p}\cos(\kappa_{i,j,n,n,l,p})+\notag\\
				&\sum_{n=1}^{N-1}\sum_{m=n+1}^{N}\sum_{l=1}^{L_i}\sum_{p=1}^{L_i}2\mu_{i,j,n,m,l,p}\cos(\kappa_{i,j,n,m,l,p}),\notag
			\end{align}	
		\end{small}where $\mu_{i,j,n,m,l,p} = |\boldsymbol{R}_j(n,m)||\boldsymbol{a}_{i}(l)||\boldsymbol{a}_{i}(p)|$ and $\kappa_{i,j,n,m,l,p} = \angle\boldsymbol{R}_j(n,m)-\angle\boldsymbol{a}_{i}(l)+\frac{2\pi}{\lambda}\rho(t_n,\psi_i^l)+\angle \boldsymbol{a}_{i}(p)-\frac{2\pi}{\lambda}\rho(t_m,\psi_i^p)$. Recalling that $\boldsymbol{t}=[t_1,t_2,\dots,t_N]^T$, the gradient vector $\nabla\tilde{\mathcal{F}}_{i,j}(\boldsymbol{t})$  w.r.t. $\boldsymbol{t}$ is given by $	\nabla\tilde{\mathcal{F}}_{i,j}(\boldsymbol{t}) = \left[ \frac{\partial \tilde{\mathcal{F}}_{i,j}(\boldsymbol{t})}{\partial t_1}, \frac{\partial \tilde{\mathcal{F}}_{i,j}(\boldsymbol{t})}{\partial t_2},...,  \frac{\partial \tilde{\mathcal{F}}_{i,j}(\boldsymbol{t})}{\partial t_N} \right]^T$,
		%		\begin{equation}
			%			\nabla\tilde{\mathcal{F}}_{i,j}(\boldsymbol{t}) = \left[ \frac{\partial \tilde{\mathcal{F}}_{i,j}(\boldsymbol{t})}{\partial t_1}, \frac{\partial \tilde{\mathcal{F}}_{i,j}(\boldsymbol{t})}{\partial t_2},...,  \frac{\partial \tilde{\mathcal{F}}_{i,j}(\boldsymbol{t})}{\partial t_N} \right]^T.
			%		\end{equation}
		%	\begin{equation}
			%	\nabla^2\tilde{\mathcal{F}}_{i,j}(\boldsymbol{t}) =	\begin{bmatrix}
				%		\begin{matrix}
					%			\frac{\partial^2 \tilde{\mathcal{F}}_{i,j}(\boldsymbol{t})}{\partial (t_1)^2} & \frac{ \partial^2 \tilde{\mathcal{F}}_{i,j}(\boldsymbol{t})}{\partial t_1 \partial t_2} \\
					%			\frac{\partial^2 \tilde{\mathcal{F}}_{i,j}(\boldsymbol{t})}{\partial t_1 \partial t_2} & \frac{\partial^2 \tilde{\mathcal{F}}_{i,j}(\boldsymbol{t})}{\partial (t_2)^2}
					%		\end{matrix}
				%		& \cdots & 
				%		\begin{matrix}
					%			\frac{\partial^2 \tilde{\mathcal{F}}_{i,j}(\boldsymbol{t})}{\partial t_1 \partial t_{N}} \\ \frac{\partial^2 \tilde{\mathcal{F}}_{i,j}(\boldsymbol{t})}{\partial t_2 \partial t_{N}}
					%		\end{matrix} \\
				%		\vdots & \ddots & \vdots \\
				%		\begin{matrix}
					%			\frac{\partial^2 \tilde{\mathcal{F}}_{i,j}(\boldsymbol{t})}{\partial t_1 \partial t_N} & \frac{\partial^2 \tilde{\mathcal{F}}_{i,j}(\boldsymbol{t})}{\partial t_2 \partial t_N}
					%		\end{matrix}
				%		& \cdots &
				%		\frac{\partial^2 \tilde{\mathcal{F}}_{i,j}(\boldsymbol{t})}{\partial (t_N)^2}
				%	\end{bmatrix}.
			%	\label{Hessian}
			%\end{equation}	
			with each element given by
			\begin{small}
				\begin{align}\label{OBJ_app}
					&\frac{\partial\tilde{\mathcal{F}}_{i,j}(\boldsymbol{t})}{\partial t_n} =\notag\\ &\sum_{l=1}^{L_i-1}\sum_{p=l+1}^{L_i}-\frac{4\pi}{\lambda}\mu_{i,j,n,n,l,p}\sin(\kappa_{i,j,n,n,l,p})\left(\cos\psi_i^l-\cos\psi_i^p\right)\nonumber\\  
					&+\sum_{m=n+1}^{N}\sum_{l=1}^{L_i}\sum_{p=1}^{L_i}-\frac{4\pi}{\lambda}\mu_{i,j,n,m,l,p}\sin(\kappa_{i,j,n,m,l,p})\cos\psi_i^l\notag\\
					&+ \sum_{m=1}^{n-1}\sum_{l=1}^{L_i}\sum_{p=1}^{L_i}\frac{4\pi}{\lambda}\mu_{i,j,m,n,l,p}\sin(\kappa_{i,j,m,n,l,p})\cos\psi^p_i.
				\end{align} 
			\end{small}
			
			Thus, the gradient vector $\nabla{\mathcal{F}}_2(\boldsymbol{t})$ can be obtained. \hfill$\Box$

			\section{Construction of approximation in~\eqref{P4_approximation}}
			We only need to show the construction of $\nabla\tilde{\Gamma}_{w,i}(\boldsymbol{t})$ and $\tilde{\delta}_{w,i}$, since the construction of $\nabla \mathcal{G}_w(\boldsymbol{t})$ and $\bar{\delta}_{w}$ can be performed in a similar fashion, and is thus omitted for brevity.  The Hessian matrix $\nabla^2\tilde{\Gamma}_{w,i}(\boldsymbol{t})$ w.r.t. $\boldsymbol{t}$ is given by
			\begin{equation}
				\nabla^2 \tilde{\Gamma}_{w,i}({\boldsymbol{t}}) =	\begin{bmatrix}
					\begin{matrix}
						\frac{\partial^2 \tilde{\Gamma}_{w,i}({\boldsymbol{t}})}{\partial (t_1)^2} & \frac{\partial^2 \tilde{\Gamma}_{w,i}({\boldsymbol{t}})}{\partial t_1 \partial t_2} \\
						\frac{\partial^2 \tilde{\Gamma}_{w,i}({\boldsymbol{t}})}{\partial t_2 \partial t_1} & \frac{\partial^2 \tilde{\Gamma}_{w,i}({\boldsymbol{t}})}{\partial (t_2)^2}
					\end{matrix}
					& \cdots & 
					\begin{matrix}
						\frac{\partial^2 \tilde{\Gamma}_{w,i}({\boldsymbol{t}})}{\partial t_1 \partial t_N} \\ \frac{\partial^2 \tilde{\Gamma}_{w,i}({\boldsymbol{t}})}{\partial t_2 \partial t_N}
					\end{matrix} \\
					\vdots & \ddots & \vdots \\
					\begin{matrix}
						\frac{\partial^2 \tilde{\Gamma}_{w,i}({\boldsymbol{t}})}{\partial t_N \partial t_1} & \frac{\partial^2 \tilde{\Gamma}_{w,i}({\boldsymbol{t}})}{\partial t_N \partial t_2}
					\end{matrix}
					& \cdots &
					\frac{\partial^2 \tilde{\Gamma}_{w,i}({\boldsymbol{t}})}{\partial (t_N)^2}
				\end{bmatrix}.
				\label{Hessian}
			\end{equation}
			 We then rewrite $\tilde{\Gamma}_{w,i}(\boldsymbol{t})$ as 
			\begin{align}
				&\tilde{\Gamma}_{w,i}(\boldsymbol{t}) = \mathbf{a}_t(\varphi_w,\boldsymbol{t})^H\boldsymbol{R}_{w,w}^i\mathbf{a}_t(\varphi_w,\boldsymbol{t})\notag\\
				&-\sum_{c\neq w}^{W}\Gamma_w\mathbf{a}_t(\varphi_c,\boldsymbol{t})^H\boldsymbol{R}_{w,c}^i\mathbf{a}_t(\varphi_c,\boldsymbol{t})-\sigma_r^2\Gamma_w\boldsymbol{u}_{w,i}^H\boldsymbol{u}_{w,i},
			\end{align}
			where $\boldsymbol{R}_{w,c}^i = |\alpha_c|^2\boldsymbol{u}_{w,i}^H\mathbf{a}_r(\varphi_c,\boldsymbol{r})\boldsymbol{R}_X^i\mathbf{a}_r(\varphi_c,\boldsymbol{r})^H\boldsymbol{u}_{w,i},\forall i \in\{0,1\},\forall w,c\in\mathcal{W}.$ We define that $\mathcal{V}(\mathbf{a},\boldsymbol{Q}) \triangleq   \mathbf{a}^H\boldsymbol{Q}\mathbf{a}.$ Recalling that $\mathbf{a}_t(\varphi_w,{\boldsymbol{t}}) = [e^{\jmath\frac{2\pi}{\lambda}\rho({t}_1,\varphi_w)},\dots,e^{\jmath\frac{2\pi}{\lambda}\rho({t}_{N},\varphi_w)}]^T$, the term  $\mathcal{V}(\mathbf{a}_t(\varphi_w,{\boldsymbol{t}}),\boldsymbol{R}_{w,w}^i)$ can be recast as 
			\begin{small}
				\begin{align}
					\mathcal{V}&(\mathbf{a}_t(\varphi_w,{\boldsymbol{t}}),\boldsymbol{R}_{w,w}^i) = \sum_{n=1}^{N}\boldsymbol{R}_{w,w}^i(n,n)+\sum_{n=1}^{N-1}\sum_{j=n+1}^{N}2|\boldsymbol{R}_{w,w}^i(n,j)|\notag\\
					&\cos(\frac{-2\pi}{\lambda}\rho({t}_n,\varphi_w)+\frac{2\pi}{\lambda}\rho({t}_j,\varphi_w)+\angle\boldsymbol{R}_{w,w}^i(n,j)).
				\end{align}
			\end{small}where $\boldsymbol{R}_{w,w}^i(n,j)$ denotes the $(n,j)$-th element of $\boldsymbol{R}_{w,w}^i$. The $n$-th element of the gradient vector $\nabla\mathcal{V}(\mathbf{a}_t(\varphi_w,{\boldsymbol{t}}),\boldsymbol{R}_{w,w}^i)$, i.e.,  $\frac{\partial\mathcal{V}(\mathbf{a}_t(\varphi_w,{\boldsymbol{t}}),\boldsymbol{R}_{w,w}^i)}{\partial t_n}$ can be calculated as 
			\begin{small}
				\begin{align*}
					&\frac{\partial\mathcal{V}(\mathbf{a}_t(\varphi_w,{\boldsymbol{t}}),\boldsymbol{R}_{w,w}^i)}{\partial t_n}=\sum_{j=n+1}^{N}\frac{4\pi}{\lambda}|\boldsymbol{R}_{w,w}^i(n,j)|\sin(\frac{-2\pi}{\lambda}\rho({t}_n,\varphi_w)\notag\\
					&+\frac{2\pi}{\lambda}\rho({t}_j,\varphi_w)+\angle \boldsymbol{R}_{w,w}^i(n,j))\cos\varphi_w+\sum_{j=1}^{n-1}\frac{-4\pi}{\lambda}|\boldsymbol{R}_{w,w}^i(j,n)|\notag\\
					&\sin(\frac{-2\pi}{\lambda}\rho({t}_j,\varphi_w)+\frac{2\pi}{\lambda}\rho({t}_n,\varphi_w)+\angle\boldsymbol{R}_{w,w}^i(j,n))\cos\varphi_w.
				\end{align*}
			\end{small}Thus, the gradient vector $\nabla\tilde{\Gamma}_{w,i}(\boldsymbol{t}) = \frac{\partial\mathcal{V}(\mathbf{a}_t(\varphi_w,{\boldsymbol{t}}),\boldsymbol{R}_{w,w}^i)}{\partial \boldsymbol{t}}-\sum_{c\neq w}^{W}\Gamma_w\frac{\partial\mathcal{V}(\mathbf{a}_t(\varphi_c,{\boldsymbol{t}}),\boldsymbol{R}_{w,c}^i)}{\partial \boldsymbol{t}}$ is obtained. To facilitate the derivation, we further define that $\beta_{n,j,w,c} = \frac{-2\pi}{\lambda}\rho(t_n,\varphi_w)+\frac{2\pi}{\lambda}\rho(t_j,\varphi_w)+\angle \boldsymbol{R}_{w,c}^i(n,j)$. Then, the calculation on elements of $\nabla^2\mathcal{V}(\mathbf{a}_t(\varphi_w,{\boldsymbol{t}}),\boldsymbol{R}_{w,w}^i)$ is given in~\eqref{Hessian_element}. With $\frac{\partial^2 \tilde{\Gamma}_{w,i}({\boldsymbol{t}})}{\partial t_n\partial t_j} = \frac{\partial^2 \mathcal{V}(\mathbf{a}_t(\varphi_w,{\boldsymbol{t}}),\boldsymbol{R}_{w,w}^i)}{\partial t_n\partial t_j}-\sum_{c\neq w}^{W}\Gamma_w\frac{\partial^2\mathcal{V}(\mathbf{a}_t(\varphi_c,{\boldsymbol{t}}),\boldsymbol{R}_{w,c}^i)}{\partial t_n\partial t_j}$, the Hessian matrix $\nabla^2 \tilde{\Gamma}_{w,i}({\boldsymbol{t}})$ can be obtained. Since $||\nabla^2 \tilde{\Gamma}_{w,i}({\boldsymbol{t}})||_2^2\le||\nabla^2 \tilde{\Gamma}_{w,i}({\boldsymbol{t}})||_F^2$ and $||\nabla^2 \tilde{\Gamma}_{w,i}({\boldsymbol{t}})||_2\boldsymbol{I}_N\succeq \nabla^2 \tilde{\Gamma}_{w,i}({\boldsymbol{t}})$, we can select $\tilde{\delta}_{w,i}$ as
			\begin{small}
			\begin{equation*}
				\tilde{\delta}_{w,i} = \frac{8N^2\pi^2}{\lambda^2} \Big(\underset{n,j}{\max}~\boldsymbol{R}_{w,w}^i(n,j)+\Gamma_w\sum_{c\neq w}^{W}\underset{n,j}{\max}~\boldsymbol{R}_{w,c}^i(n,j)\Big).
			\end{equation*}
			\end{small}
			\vspace{-5mm}
					\begin{figure*}[tp]
						\begin{small}
							\begin{subequations}\label{Hessian_element}	
								\begin{alignat}{2}
									&\frac{\partial^2\mathcal{V}(\mathbf{a}_t(\varphi_w,{\boldsymbol{t}}),\boldsymbol{R}_{w,w}^i)}{\partial(t_n)^2} = \sum_{j=n+1}^{N}\frac{-8\pi^2}{\lambda^2}|\boldsymbol{R}_{w,w}^i(n,j)|\cos\beta_{n,j,w,w}\cos^2\varphi_w +\sum_{j=1}^{n-1}\frac{-8\pi^2}{\lambda^2}|\boldsymbol{R}_{w,w}^i(j,n)|\cos\beta_{j,n,w,w}\cos^2\varphi_w,\\ 
									&\frac{\partial^2\mathcal{V}(\mathbf{a}_t(\varphi_w,{\boldsymbol{t}}),\boldsymbol{R}_{w,w}^i)}{\partial t_n \partial t_j}=	
									\begin{cases} 
										&\frac{8\pi^2}{\lambda^2}|\boldsymbol{R}_{w,w}^i(n,j)|\cos\beta_{n,j,w,w}\cos^2\varphi_w,j > n,\\ 
										&\frac{8\pi^2}{\lambda^2}|\boldsymbol{R}_{w,w}^i(j,n)|\cos\beta_{j,n,w,w}\cos^2\varphi_w,j < n,
									\end{cases} 
								\end{alignat} 
							\end{subequations} 
						\end{small}
					\hrule
					\vspace{-5mm}
				\end{figure*}

			%				\begin{figure*}[htbp]
				%					\begin{small}
					%					\begin{align}\label{OBJ_app}
						%						&\frac{\partial\tilde{\mathcal{F}}_{i,j}(\boldsymbol{t})}{\partial t_n} = \sum_{l=1}^{L_i-1}\sum_{p=l+1}^{L_i}-\frac{4\pi}{\lambda}\mu_{i,j,n,n,l,p}\sin(\kappa_{i,j,n,n,l,p})\left(\cos\psi_i^l-\cos\psi_i^p\right)+ \nonumber\\ 
						%						&\sum_{m=n+1}^{N}\sum_{l=1}^{L_i}\sum_{p=1}^{L_i}-\frac{4\pi}{\lambda}\mu_{i,j,n,m,l,p}\sin(\kappa_{i,j,n,m,l,p})\cos\psi_i^l+ \sum_{m=1}^{n-1}\sum_{l=1}^{L_i}\sum_{p=1}^{L_i}\frac{4\pi}{\lambda}\mu_{i,j,m,n,l,p}\sin(\kappa_{i,j,m,n,l,p})\cos\psi^p_i.
						%					\end{align} 
					%					\end{small}	
				%					\vspace{-2mm}	
				%					\hrule
				%					\vspace{-5mm}
				%				\end{figure*}

			%	\begin{figure}[t]
				%		\centering
				%		\includegraphics[width=0.4\textwidth]{antenna.eps}
				%		\captionsetup{font={normalsize},labelsep=period,singlelinecheck=off}
				%		\caption{The radar SINR versus antenna number.} 
				%		\label{antenna} 
				%	\end{figure}%

			\vspace{-4mm}

			\balance

			\bibliography{reference.bib} 
			\bibliographystyle{IEEEtran} 
			
			% \bibliography{reference}
			
			\vfill
		\end{document}